\documentclass[
    reprint,
    twocolumn,
    noeprint,
    superscriptaddress,
nofootinbib,
    floatfix
    ]{revtex4-2}

\usepackage[p]{newtxtext}
\usepackage[vvarbb,upint]{newtxmath}
\usepackage[mathcal]{euscript}
\usepackage[varl]{zi4}
\usepackage{microtype}

\usepackage{verbatim}

\newcommand{\rmd}{\mathrm{d}}
\newcommand{\rme}{\mathrm{e}}

\newcommand{\mr}[1]{\mathrm{#1}}
\DeclareMathOperator{\im}{Im}

\usepackage{amsmath}
\usepackage[
    range-units=single,
]{siunitx}
\usepackage{graphicx}
\usepackage[dvipsnames,x11names]{xcolor}
\usepackage[colorlinks=True,
            linkcolor=DeepPink4,
            citecolor=DodgerBlue4,
            urlcolor=PaleGreen4]{hyperref}

\usepackage[capitalise]{cleveref}

                                      \providecommand\apj{ApJ}                 \providecommand\apjl{ApJ}                               \providecommand\ao{Appl.~Opt.}                                                                                 \providecommand\baas{BAAS}

                                                                        \providecommand\prb{Phys.~Rev.~B}                \providecommand\prd{Phys.~Rev.~D}                \providecommand\prl{Phys.~Rev.~Lett.}    \providecommand\pasa{PASA}                                                                                                                               
 
     \providecommand\grl{Geophys.~Res.~Lett.}           

   \providecommand\physrep{Phys.~Rep.}

 \newread\gwincfile
\openin\gwincfile=data/CEparams.txt
\def\DefGwincVal#1: #2\relax{\expandafter\gdef\csname gwincval-#1\endcsname{#2}}
{\endlinechar=-1
\loop
\ifeof\gwincfile\else
\read\gwincfile to \tmp
\ifx\tmp\empty\else
\expandafter\DefGwincVal\tmp\relax
\fi
\repeat
}
\closein\gwincfile
\providecommand{\GwincVal}[1]{\csname gwincval-#1\endcsname}

\usepackage{xparse}
\ExplSyntaxOn
  \cs_new_eq:NN \fpeval \fp_eval:n
\ExplSyntaxOff
 \newcommand\aligoBnsHorizon{0.097}
\newcommand\aligoBbhHorizon{1.21}
\newcommand\aligoMaxMtot{710}
 \newcommand\aligoEarlyWarning{12.4\,\si{\second}} \newcommand\voyBnsHorizon{0.44}
\newcommand\voyBbhHorizon{7.4}
\newcommand\voyMaxMtot{950}
 \newcommand\voyEarlyWarning{4.2\,\si{\minute}} \newcommand\etBnsHorizon{3.6}
\newcommand\etBbhHorizon{57}
\newcommand\etMaxMtot{4e+03}
 \newcommand\etEarlyWarning{5.3\,\si{\hour}} \newcommand\ceOneBnsHorizon{4.1}
\newcommand\ceOneBbhHorizon{34}
\newcommand\ceOneMaxMtot{2.2e+03}
 \newcommand\ceOneEarlyWarning{65\,\si{\minute}} \newcommand\ceTwoSiliconBnsHorizon{11.3}
\newcommand\ceTwoSiliconBbhHorizon{41}
\newcommand\ceTwoSiliconMaxMtot{2.4e+03}
 \newcommand\ceTwoSiliconEarlyWarning{90\,\si{\minute}}

\DeclareSIUnit{\torr}{torr}

\newcommand{\figwidth}{\columnwidth}

\newcommand{\LigoMIT}{LIGO Laboratory, Massachusetts Institute of Technology, Cambridge, Massachusetts 02139, USA}
\newcommand{\CSUFullerton}{Nicholas and Lee Begovich Center for Gravitational-Wave Physics and Astronomy, California State University, Fullerton, Fullerton, California 92831, USA}
\newcommand{\LigoCaltech}{LIGO Laboratory, California Institute of Technology, Pasadena, California 91125, USA}
\newcommand{\BurkeCaltech}{Walter Burke Institute for Theoretical Physics, California Institute of Technology, Pasadena, California 91125, USA}
\newcommand{\Syracuse}{Department of Physics, Syracuse University, Syracuse, New York 13244, USA}
\newcommand{\GSSI}{Gran Sasso Science Institute (GSSI), I-67100 L'Aquila, Italy}
\newcommand{\GranSasso}{INFN, Laboratori Nazionali del Gran Sasso, I-67100 Assergi, Italy}
\begin{document}
\newcommand{\OzGravANU}{OzGrav, ANU Centre for Gravitational Astrophysics, Research Schools of
Physics, and Astronomy and Astrophysics, The Australian National University, Canberra, ACT 2601, Australia}
\newcommand{\UCSC}{Department of Astronomy and Astrophysics, University of California Santa  Cruz, Santa Cruz, California 95064, USA}
\newcommand{\MechECaltech}{Department of Mechanical and Civil Engineering, California Institute of Technology, Pasadena, California 91125, USA}

\title[Cosmic Explorer low frequency]{Gravitational-wave physics with Cosmic Explorer: limits to low-frequency sensitivity}

\author{Evan D. Hall}
\address{\LigoMIT}
\author{Kevin Kuns}
\address{\LigoMIT}
\author{Joshua R. Smith}
\address{\CSUFullerton}
\author{Yuntao Bai}
\address{\BurkeCaltech}
\author{Christopher Wipf}
\address{\LigoCaltech}
\author{Sebastien Biscans}
\address{\LigoCaltech}
\address{\LigoMIT}
\author{Rana X Adhikari}
\address{\LigoCaltech}
\author{Koji Arai}
\address{\LigoCaltech}
\author{Stefan Ballmer}
\address{\Syracuse}
\author{Lisa Barsotti}
\address{\LigoMIT}
\author{Yanbei Chen}
\address{\BurkeCaltech}
\author{Matthew Evans}
\address{\LigoMIT}
\author{Peter Fritschel}
\address{\LigoMIT}
\author{Jan Harms}
\address{\GSSI}
\address{\GranSasso}
\author{Brittany Kamai}
\address{\UCSC}
\address{\MechECaltech}
\author{Jameson Graef Rollins}
\address{\LigoCaltech}
\author{David Shoemaker}
\address{\LigoMIT}
\author{Bram Slagmolen}
\address{\OzGravANU}
\author{Rainer Weiss}
\address{\LigoMIT}
\author{Hiro Yamamoto}
\address{\LigoCaltech}

\begin{abstract}
Cosmic Explorer (CE) is a next-generation ground-based gravitational-wave
observatory concept, envisioned to begin operation in the 2030s, and expected
to be capable of observing binary neutron star and black hole mergers back to
the time of the first stars.  Cosmic Explorer's sensitive band will extend
below \SI{10}{\Hz}, where the design is predominantly limited by geophysical, thermal, and quantum  noises.  In this work, thermal, seismic, gravity-gradient, quantum, residual gas,
scattered-light, and servo-control noises are analyzed in order to motivate
facility and vacuum system design requirements, potential test mass suspensions, Newtonian noise
reduction strategies, improved inertial sensors, and cryogenic control
requirements.  Our analysis shows that with improved technologies, Cosmic
Explorer can deliver a strain sensitivity better than \SI{e-23}{\Hz^{-1/2}}
down to \SI{5}{\Hz}.  Our work refines and extends previous analysis of the
Cosmic Explorer concept and outlines the key research areas
needed to make this observatory a reality.
\end{abstract}

\maketitle

\section{Introduction}
\label{sec:introduction}

The second-generation of laser interferometric gravitational-wave
observatories\,---\,Advanced LIGO~\cite{2015CQGra..32g4001L}, Advanced
Virgo~\cite{2015CQGra..32b4001A}, and
Kagra~\cite{2019NatAs...3...35K}\,---\,have opened a new window on the universe
by observing gravitational waves from merging systems of black
holes~\cite{2016PhRvL.116f1102A,2016PhRvX...6d1015A} and neutron
stars~\cite{2017PhRvL.119p1101A}, and have ushered in a new era in
multi-messenger astronomy~\cite{2017ApJ...848L..12A}.  Dozens of coalescing
binary systems have been observed thus
far~\cite{2019PhRvX...9c1040A,2019ApJ...872..195N}, with rapid alerts
delivering sky locations and probable system types~\cite{2019ApJ...875..161A},
bringing the features of the underlying astrophysical populations into focus.
An enhancement to Advanced LIGO, known as LIGO A+, with improved quantum noise
and optical coatings, is now being implemented~\cite{2015PhRvD..91f2005M}.
Additionally, research and development is underway toward a cryogenic silicon
detector, LIGO Voyager, that could be implemented in the existing LIGO
facilities~\cite{2020arXiv200111173A}, and a concept for a
high-frequency-focused Australian observatory is being
developed~\cite{2020PASA...37...47A}.

A vision is developing for a global third-generation (3G) network of
ground-based gravitational-wave observatories capable of observing
gravitational waves across cosmic time, with nearby systems detected with
incredible precision~\cite{gwic3gsc, gwic3grd}.  The European concept for a
third-generation observatory is the Einstein Telescope
(ET)~\cite{2011CQGra..28i4013H}, a \SI{10}{\km} triangular underground
observatory combining three high-power room-temperature interferometers
sensitive at high frequency and three cryogenic silicon interferometers
sensitive at low frequency.  A United States concept for a 3G observatory is
Cosmic
Explorer~\cite{2017CQGra..34d4001A,2019BAAS...51c.141R,2019BAAS...51g..35R}, a
\SI{40}{\kilo\m} L-shaped, single-interferometer observatory built on the
Earth's surface.

We anticipate a staged approach to Cosmic Explorer, similar to the approach
adopted by LIGO, in which the facility hosts successive generations of
detectors, each exploiting the most advanced technology available at the time.
We envision that the first detector, Cosmic Explorer 1 (CE1), will operate in
the 2030s using LIGO A+ technology scaled up to the increased dimensions of the
facility, and with a few modest improvements.  For the 2040s, the
state-of-the-art technology is more difficult to predict.  In this work we
consider two possible designs for this detector, called Cosmic Explorer 2
(CE2).  One possibility is that CE2 is a further extension of LIGO A+
technology, retaining room-temperature fused silica test masses and a
\SI{1}{\um} laser as the working technology.  Another possibility is that CE2
is an extension of the LIGO Voyager technology, employing silicon test masses
at \SI{123}{\kelvin} and a \SI{2}{\um} laser. In the rest of the paper, we will
refer to detectors based on room temperature fused silica test masses and
\SI{1}{\um} laser wavelength as the ``\SI{1}{\um} technology'' and those with
cryogenic silicon test masses and \SI{2}{\um} laser wavelength as the
``\SI{2}{\um} technology.'' For both CE1 and CE2, the detector designs target
observations above \SI{5}{\Hz}, while Einstein Telescope targets observations
down to \SI{3}{\Hz}.

This paper presents an assessment of the low-frequency sensitivity of CE1 and
CE2 based on recent research and development progress.  We first present the
basic low-frequency observational capabilities of Cosmic Explorer
(\cref{sec:astrophysics}) and discuss broadly the limits to the Cosmic Explorer
strain sensitivities (\cref{sec:limits}).  We then describe the Cosmic Explorer
facility (\cref{sec:facility}) and go into detail about low-frequency noise
sources (\cref{sec:noise}).  Then in \cref{sec:discussion} we take stock of the
research and development that will be required to realize Cosmic Explorer and
we look forward to future work. \cref{sec:technologies} summarizes the different technologies used in the two stages of Cosmic Explorer and \cref{sec:displacement_force} compares the displacement and force noises of Cosmic Explorer with those of other detectors.
 
\section{Astrophysics}
\label{sec:astrophysics}

\begin{figure}[t]
    \centering
    \includegraphics[width=\figwidth]{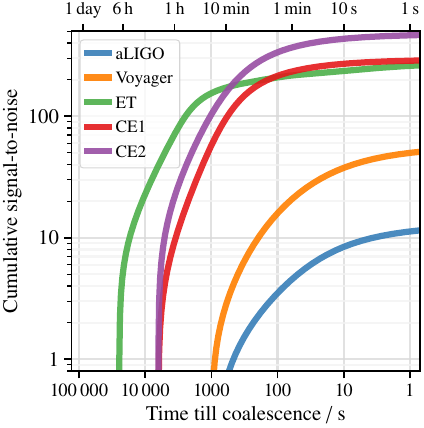}
    \caption{
    Signal-to-noise ratio (SNR) accumulation of a 1.4+1.4$M_\odot$ binary neutron
star system at redshift $z = 0.03$, optimally oriented.  The low-frequency
cutoffs are the same as given in \cref{fig:gw_strains}. Numerical early warning
values for a threshold signal-to-noise ratio of 8 are given in
\cref{tab:astro}, showing that third-generation detectors will provide early
warning on the scale of hours, compared to the minutes provided by
second-generation detectors.  Systems this loud (or louder) should be expected
roughly once per year assuming a local merger rate of
${\sim}\SI{300}{Gpc^{-3}\,yr^{-1}}$~\cite{2020arXiv201014533T}.
\label{fig:early_warning}}
\end{figure}

\begin{figure}[t]
    \centering
    \includegraphics[width=\figwidth]{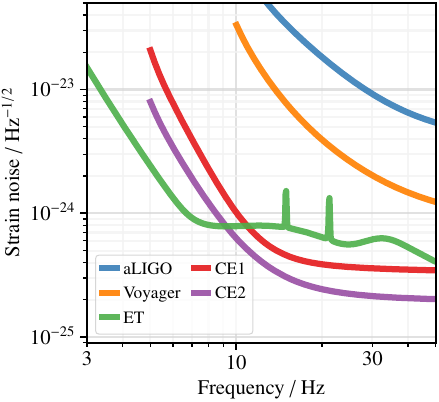}
    \caption{
    Strain noise of Advanced LIGO, LIGO Voyager, the six-interferometer
Einstein Telescope, and both stages of Cosmic Explorer.  In all cases the
source is assumed to be circularly polarized.  aLIGO and Voyager are shown for
$f \ge \SI{10}{\Hz}$, CE for $f \ge \SI{5}{\Hz}$, and ET for $f \ge
\SI{3}{\Hz}$; these are the low-frequency cutoffs assumed for the signal
calculations throughout this work.
    }
    \label{fig:gw_strains}
\end{figure}

\begin{figure}[t]
    \centering
    \includegraphics[width=\figwidth]{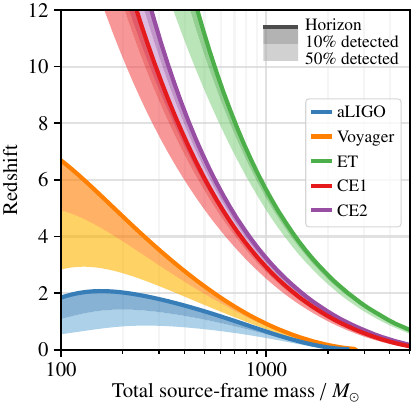}
    \caption{
    Detectability of nonspinning equal-mass
black hole binaries as a function of mass and redshift, with detectability
being defined as having a matched-filter signal-to-noise ratio (SNR) $\ge 8$.
The solid line indicates each detector's horizon, at which an optimally
oriented system with a given mass and redshift will be detected with
$\text{SNR} = 8$, and suboptimally oriented systems have $\text{SNR} < 8$.
Systems lying above the solid line are limited to $\text{SNR} < 8$ regardless
of orientation.  Along the edge of the dark (light) shaded band, \SI{10}{\%}
(\SI{50}{\%}) of the systems will be detected with $\text{SNR} \ge 8$ and the
remainder will have $\text{SNR} < 8$ due to unfavorable orientation.
    }
    \label{fig:gw_horizons_seeds}
\end{figure}

\begin{table}[t]
    \centering
\begin{tabular}{
        r
        S[round-mode=figures,round-precision=2,scientific-notation=fixed]
        S[round-mode=places, round-precision=0,scientific-notation=fixed,table-align-text-post = false]
        S[round-mode=figures,round-precision=2,scientific-notation=fixed]
        S[round-mode=figures,round-precision=2,scientific-notation=fixed]}
        \toprule
        Observatory &
            {$z_\text{hor}^{\text{(BNS)}}$} &
            {$t_\text{early}^{\text{(BNS)}}$} &
            {$z_\text{hor}^{\text{(BBH)}}$} &
            {$M_\text{max}$ / $M_\odot$} \\
        \hline
        aLIGO   &
            \aligoBnsHorizon{}   &
            \aligoEarlyWarning{} &
            \aligoBbhHorizon{}   &
            \aligoMaxMtot{} \\
        Voyager &
            \voyBnsHorizon{} &
            \voyEarlyWarning{} &
            \voyBbhHorizon{} &
            \voyMaxMtot{} \\
        CE1     &
            \ceOneBnsHorizon{}   &
            \ceOneEarlyWarning{} &
            \ceOneBbhHorizon{}   &
            \ceOneMaxMtot{} \\
        CE2     &
            \ceTwoSiliconBnsHorizon{}   &
            \ceTwoSiliconEarlyWarning{} &
            \ceTwoSiliconBbhHorizon{}   &
            \ceTwoSiliconMaxMtot{} \\
        ET      &
            \etBnsHorizon{}      &
            \etEarlyWarning{} &
            \etBbhHorizon{}      &
            \etMaxMtot{} \\
        \botrule
    \end{tabular}
\caption{
    Observational performance metrics for the sensitivities shown in
\cref{fig:gw_strains}. ``BNS'' refers to a 1.4+1.4$M_\odot$ neutron-star system
(tidal and post-merger effects not included), and ``BBH'' to a 30+30$M_\odot$
black hole system, in both cases nonspinning.  The time before merger is given
for an optimally oriented neutron star system at a redshift $z = 0.03$, with a
threshold signal-to-noise of 8.  $M_\text{max}$ is the maximum mass for which
an optimally oriented nonspinning equal-mass system could be detected at $z =
1$.
    }
    \label{tab:astro}
\end{table}

Cosmic Explorer has a range of science goals, which together take advantage of
the instrument's full broadband sensitivity up to several kilohertz.  Here we
focus on the detection of compact binary signals.  The low-frequency
sensitivity affects the reach of the instrument for heavy and high-redshift
signals, as well as the total signal-to-noise ratio of all compact-binary
signals.  The optimal signal-to-noise ratio for a particular
frequency-domain signal $h(f)$, measured in a detector with a strain sensitivity $S_h$ and a gravitational-wave sensitivity band extending down to a low-frequency cutoff $f_\text{low}$, is obtained when the signal is detected with a matched filter, yielding an amplitude signal-to-noise ratio $\rho$ given by $\rho^2 = 4\int_{f_\text{low}}^\infty \!\rmd\!f |h(f)|^2/S_h(f)$~\cite{1970esn..book.....W,1993PhRvD..47.2198F,2011gwpa.book.....C}.

For light systems (e.g., neutron stars) which are still in their
inspiral phase at frequencies $f \lesssim \SI{10}{\Hz}$, improving the
low-frequency cutoff $f_\text{low}$ has a modest but noticeable improvement on
the total signal-to-noise ratio: for the idealized case of a detector
with a flat noise floor down to a lower cutoff frequency $f_\text{low}$, the
matched-filter signal-to-noise ratio scales as $\rho \propto
f_\text{low}^{-2/3}$, since $|h(f)| \propto f^{-7/6}$ in the stationary phase
approximation~\cite{1994PhRvD..49.2658C}.  This scaling amounts to roughly a
\SI{60}{\%} improvement as the cutoff frequency is halved.  On the other hand,
the improvement in the amount of early warning for these inspiraling systems
can be significant: the time $t_\text{merge}$ before a coalescing system merges
is related (again in the stationary-phase approximation) to the
gravitational-wave frequency $f_\text{GW}$ by $t_\text{merge} \propto
f_\text{GW}^{-8/3}$~\cite{2011gwpa.book.....C}. Therefore, sufficiently loud signals will accumulate threshold signal-to-noise ratio soon
after entering the sensitivity band, leading to an early warning time $t_\text{early}
\propto f_\text{low}^{-8/3}$.  This means that halving the low-frequency cutoff
increases the early warning time more than sixfold.  \cref{fig:early_warning}
shows the accumulation of signal-to-noise ratio $\rho$ found by computing
$\rho^2(t_\text{merge}) = 4\int_{f_\text{low}}^{f_\text{GW}(t_\text{merge})} \! \rmd\!f
|h(f)|^2 / S_h(f)$, with $h(f)$ in this case chosen to correspond with a
1.4+1.4$M_\odot$ binary system at $z = 0.03$ (luminosity distance
\SI{0.14}{Gpc}). By setting a threshold SNR of 8, the corresponding early
warning time can be solved for, and is given in \cref{tab:astro} for the
detector sensitivities shown in \cref{fig:gw_strains}.

Low-frequency sensitivity is especially impactful for the detection of
intermediate-mass black holes in the range $100\,M_\odot \lesssim M \lesssim
1000\,M_\odot$.  Detecting these systems at redshifts approaching 10 would
provide information on the oldest population of stars (Population III).
Additionally, these detections could demonstrate that supermassive black
holes---approaching and exceeding $10^6 M_\odot$---were formed by accretion and
hierarchical mergers from Population III remnants (the so-called ``light seed''
scenario)~\cite{2009ApJ...698L.129S}.  \cref{fig:gw_horizons_seeds} shows the
response distance~\cite{2021CQGra..38e5010C}\,---\,the redshift out to which
binary black hole systems can be detected\,---\,for Cosmic Explorer and other
detectors.  Computing the response distance for a threshold signal-to-noise
ratio $\rho_0$ requires numerically solving $\rho_0^2 = 4\int_{f_\text{low}}^\infty \!
\rmd\!f \, |h(f;z)|^2 / S_h(f)$ for the corresponding threshold $z$; here
$h(f;z)$ is the redshifted (i.e., detector-frame) gravitational waveform, which
is obtained from the source-frame waveform $h(f)$ by the substitutions\footnote{These scalings are valid if the detector does not significantly move, due to the earth's rotation and orbit around the sun, while the signal is in that detector's sensitivity band. This approximation is valid for Cosmic Explorer, since $f_\text{low}$ is sufficiently high, but not for spaced-based detectors such as LISA.}
$f \mapsto f/(1+z)$, $m_1 \mapsto (1+z)m_1$, and $m_2 \mapsto
(1+z)m_2$~\cite{1993PhRvD..48.4738M}.  \cref{tab:astro} summarizes the
detection prospects for high-redshift, intermediate-mass black hole mergers.
 
\section{Strain sensitivity}
\label{sec:limits}

\begin{figure}[t]
  \centering
  \includegraphics[width=\figwidth]{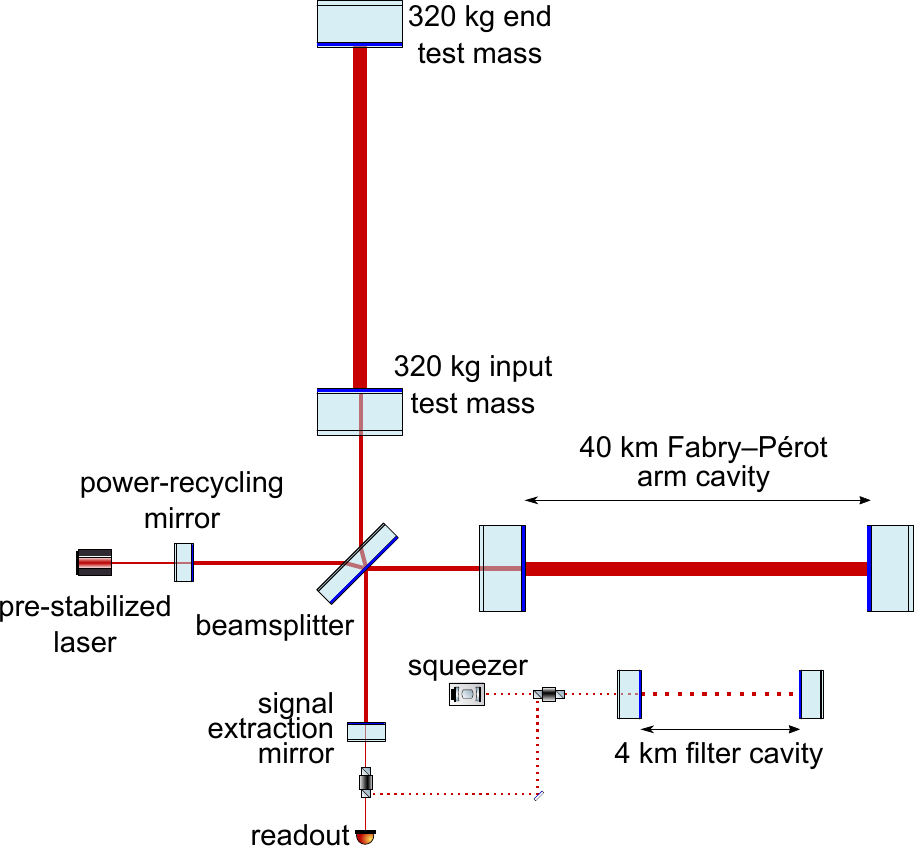}
  \caption{Simplified Cosmic Explorer interferometer topology consisting of a dual-recycled Fabry--P\'{e}rot Michelson interferometer in addition to a squeezer and filter cavity used to achieve broadband quantum noise reduction.}
  \label{fig:drfpmi}
\end{figure}

\begin{figure}[t]
  \centering
  \includegraphics[width=\figwidth]{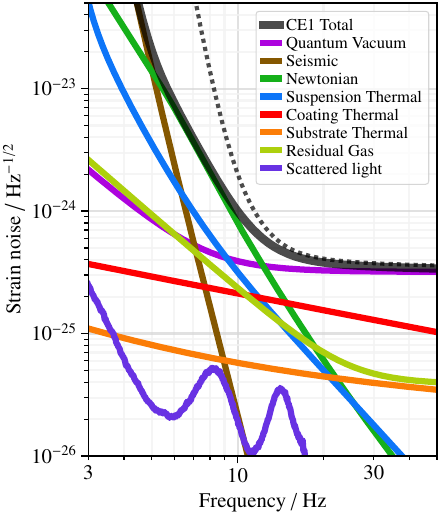}
  \caption{
    Estimated low-frequency spectral sensitivity limit (solid black) of Cosmic
Explorer 1 and the known noise sources that cause these limits (colored
curves).  The sensitivity limit from previous work~\cite{2017CQGra..34d4001A}
is also shown (dotted black curve).  From \SIrange{5}{10}{\Hz}, the strain
sensitivity is limited by seismic Newtonian noise.
    }
  \label{fig:CE1}
\end{figure}

\begin{figure*}[t]
    \centering
\includegraphics[width=\figwidth]{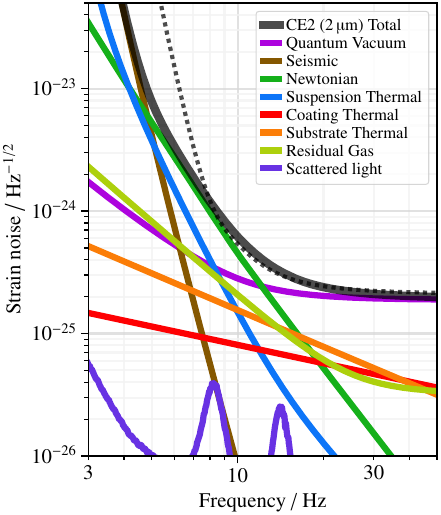}
\includegraphics[width=\figwidth]{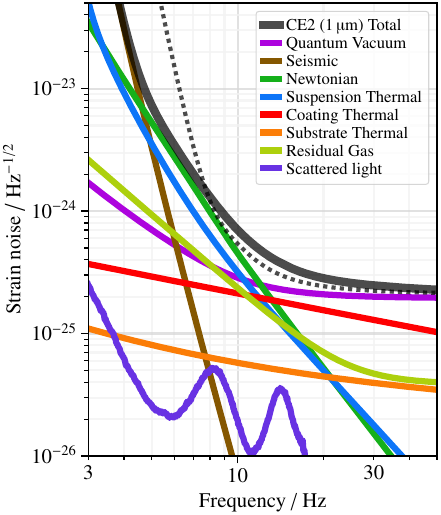}
    \caption{
Same as \cref{fig:CE1} but for Cosmic Explorer 2 realized with
\textit{Left:} the \SI{2}{\um} technology (cryogenic silicon test masses and a
\SI{2}{\um} laser wavelength) and \textit{Right:} the \SI{1}{\um} technology
(room temperature fused silica test masses and a \SI{1}{\um} laser).  For both
technologies, the seismic and suspension thermal noises are comparable to the infrasonic Newtonian noise background, which is taken to be a geophysical limit for the facility (\cref{subsubsec:newtonian_atmospheric}).
    }
    \label{fig:CE2}
\end{figure*}

Our Cosmic Explorer models adopt the dual-recycled Fabry--P\'{e}rot Michelson
interferometer topology now employed by advanced detectors shown in \cref{fig:drfpmi}.  In brief, these
detectors are Michelson interferometers whose arms are enhanced by the
inclusion of partially transmissive input mirrors, turning the arms into
Fabry--P\'{e}rot cavities. Then, a power-recycling mirror is placed between the
laser and the beamsplitter to critically couple the Fabry--P\'{e}rot arms to
the laser, which maximizes the circulating arm power. Additionally, a signal
extraction mirror between the beamsplitter and the output port is used to broaden
the bandwidth of the instrument~\cite{2017CQGra..34a5001I}. Squeezed vacuum states are reflected off of a filter cavity and injected into the antisymmetric port of the interferometer in order to achieve broadband quantum noise reduction.

The upper limit to the achievable bandwidth of Cosmic Explorer is defined by
the free spectral range of the $L = \SI{40}{\km}$ arms, given by $f_\text{FSR}
= c/2L = \SI{3.75}{\kHz}$. We take the lower limit to be \SI{5}{\Hz}; this is
not a precisely motivated cutoff, but comes from our expectation of significant
noise from local gravity fluctuations at a few hertz from the atmosphere, and
(if Advanced-LIGO-like suspensions are used) from thermal, seismic, and control
noise.  The rest of the present work is concerned primarily with the
geophysical and thermal noises, leaving a detailed discussion of other noises
to later works.

Since the initial exploration of the Cosmic Explorer
sensitivity~\cite{2017CQGra..34d4001A}, many of the estimates of the
fundamental noises have been refined, and some new noise sources have been
considered. \cref{fig:CE1,fig:CE2} show the updated low-frequency limits to the spectral
sensitivity for CE1 and CE2, respectively, and some of the key sources of noise
that contribute to these limits; the curves from the previous sensitivity study
are also included. For CE1, updates with respect to previous work mean that the
instrument attains strain noise better than \SI{e-23}{\Hz^{-1/2}} above about
\SI{5.7}{\Hz}, whereas the instrument presented in previous work attained this
performance only above \SI{8}{\Hz}.  For CE2, strain noise below
\SI{e-23}{\Hz^{-1/2}} is achieved around \SI{4.8}{\Hz} compared to \SI{6.3}{\Hz} in
previous work; additionally, the noise performance around \SI{10}{\Hz} is
slightly degraded for CE2.
The primary differences from this initial work are as follows.
\begin{itemize}

    \item\label{item:ground_motion} The ground motion of the Cosmic Explorer
facility is assumed to be lower than the LIGO facilities above \SI{5}{\Hz},
based on long-term seismic surveys from some promising locations around the
United States (\cref{sec:facility}).  This lowers both the seismic noise, and
the seismic component of the Rayleigh-wave Newtonian noise.

    \item CE1 assumes tenfold better seismic isolation than Advanced LIGO at
\SI{1}{\Hz}, and CE2 assumes one hundredfold better seismic isolation than
Advanced LIGO at \SI{1}{\Hz} (\cref{subsec:seismic}).

    \item The Newtonian noise estimates now include contributions from seismic
body waves and atmospheric infrasound (\cref{subsec:newtonian}), and CE1
assumes twofold suppression of ambient Rayleigh waves.  Together with the
reassessment of the ground motion, we find that suppression of Rayleigh and
body waves is needed for CE2 to meet the sensitivity quoted
in~\cite{2017CQGra..34d4001A}.

    \item Phase noise induced by light propagation in the bulk of the input
test masses is now included; this constitutes a potentially non-negligible noise source for
the CE2 \SI{2}{\um} technology (\cref{subsec:TMthermal}).

    \item The force noise caused by the residual gas molecules in the test mass chambers striking the test masses is now included (\cref{subsec:resgas}).

    \item The possibility of building a room temperature CE2 with \SI{1}{\um}
technology was not previously considered.  Such a detector would have
non-negligible coating thermal noise around \SI{10}{\Hz} and thus slightly
worse performance than the \SI{2}{\um} technology and the estimate from
previous work at these frequencies (\cref{subsec:TMthermal}).

    \item The suspensions for both detectors have been enlarged to \SI{4}{\meter}
of total height (previously they were \SI{3.2}{\meter}) and \SI{1500}{\kg} of
total mass (previously they were \SI{980}{\kg}), and optimized for minimal
thermal and seismic noise given updated mechanical constraints on the strength
of the materials (\cref{subsec:susthermal}).

    \item Preliminary considerations of the scattered light noise
(\cref{subsec:scatter}) and control system noise (\cref{subsec:controls})
suggest that these noises can be rendered subdominant within the
gravitational-wave band.

\end{itemize}

\section{The Cosmic Explorer facility}
\label{sec:facility}

Many of the limits to Cosmic Explorer sensitivity at low frequency depend on assumptions about the Cosmic Explorer facility and environment.
In this section we lay down requirements for the ground motion and seismic wave content (\cref{subsec:facility_seismic}), the atmospheric infrasound spectrum (\cref{subsec:facility_atmospheric}), and the ultra-high vacuum system (\cref{subsec:facility_tube}).
This list is not exhaustive; for example, magnetic requirements are not discussed because the coupling of local magnetic fields depends primarily on technical details of the detector's electronics, which are difficult to estimate without detailed modeling.
Similar topics are being considered for the underground Einstein Telescope facility~\cite{2020RScI...91i4504A}.

\subsection{Ground motion}
\label{subsec:facility_seismic}

\begin{figure}[t]
    \centering
    \includegraphics[width=\figwidth]{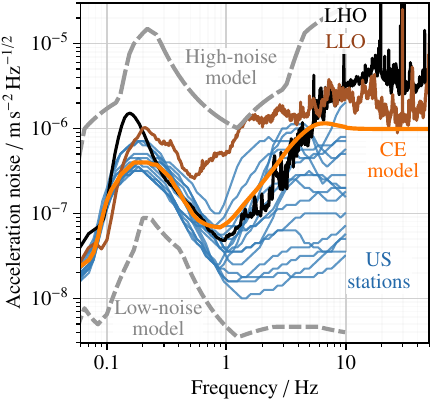}
    \caption{
        Model for Cosmic Explorer ground motion, along with representative data from LIGO Hanford (LHO), LIGO Livingston (LLO), and multi-year data from selected seismic stations in the United States.
        The Peterson high- and low-noise seismic models are also shown~\cite{Peterson1993}.
        }
    \label{fig:ground_spectra}
\end{figure}

Ground motion limits the performance of gravitational-wave interferometers both through the mechanical coupling from the ground to the suspension point of the test mass and through the direct gravitational attraction of the ground on the test mass (the so-called ``Newtonian noise'')~\cite{2019LRR....22....6H}.
Additionally, ground motion transferred to the beam tube can cause noise from stray light.

The location of Cosmic Explorer is not yet known, but an assumption for the local ground seismicity can be made based on publicly available seismic data and on the noise environment from existing facilities.
To get long-term trends that encompass diurnal and seasonal variations in ground motion, we examined noise histograms from selected USArray~\cite{Meltzer1999} and ANSS~\cite{Benz2001} seismic stations in the western United States; these stations were chosen for their proximity to promising Cosmic Explorer candidate sites which have favorable topographic properties.
We also examined noise histograms from the LIGO Hanford and Livingston sites.
Above a few hertz, the ground motion of the LIGO sites is dominated by on-site machinery. In particular, heating, ventilation and air conditioning systems dominate from \SIrange{1}{10}{\Hz}~\cite{2021arXiv210109935N}. We assume that it will be possible to design the Cosmic Explorer infrastructure to better isolate the interferometer from such machinery by moving the vibration sources out of the experimental buildings, putting them on dampers or on pedestals mounted separately and deeply into the ground.
The Cosmic Explorer ground noise model is shown in \cref{fig:ground_spectra}; this model assumes that above \SI{5}{\Hz}, the ground acceleration noise is no more than \SI{1}{\um\,\second^{-2}\,\Hz^{-1/2}}.

A complete estimate of the Newtonian noise requires a model of the seismic wave amplitude spectra and an understanding of their propagation through the ground.
In general, surface seismic motion is usually assumed to be dominated by surface waves (Rayleigh and Love waves) as opposed to body waves (P and S waves), although the actual composition depends on the particular site and may additionally include higher-order surface waves~\cite{BonnefoyClaudet2006}.
Because the Cosmic Explorer site is not known, we adopt a model in which the site is Rayleigh-wave dominated above \SI{5}{\Hz}, with a flat body-wave spectrum of amplitude \SI{0.3}{\um\,\second^{-1}\,\Hz^{-1/2}} composed equally of P waves, vertically polarized S waves, and horizontally polarized S waves.\footnote{Love waves are not considered because they do not occur in a homogeneous and isotropic elastic half-space; moreover, Love waves do not produce Newtonian noise because their motion is a horizontal shear.}
Newtonian noise is generated from only the Rayleigh, P, and vertically polarized S waves, because these waves either cause a vertical displacement of the ground surface or density fluctuations of the bulk.
The P-, S-, and Rayleigh-wave speeds are assumed to be $c_\text{P} = \SI{600}{\meter/\second}$, $c_\text{S} = \SI{300}{\meter/\second}$, $c_\text{R} = \SI{250}{\meter/\second}$, respectively.
These parameters, and the assumptions on the wave content of the ground motion, will have to be revised once the future Cosmic Explorer site is selected and characterized.

\subsection{Atmospheric fluctuations}
\label{subsec:facility_atmospheric}

Newtonian noise from density fluctuations in the atmosphere is likely to impact the strain sensitivity of third-generation detectors.
For Cosmic Explorer, the relevant mechanism is expected to be the propagation of infrasound (sound at frequency $f \lesssim \SI{20}{\Hz}$) in the vicinity of the test masses.
Global infrasound surveys provide noise histograms up to slightly below \SI{10}{\Hz}~\cite{2005GeoRL..32.9803B}; based on the median noise model, we take the outdoor infrasound spectrum for Cosmic Explorer to be \SI{1}{\milli\pascal\,\Hz^{-1/2}}.
The choice of the median infrasound background means that, while it may be possible to find a site with lower infrasound background, we are not reliant on finding an exceptional site in order to realize the noise performance described herein.
The impact of atmospheric infrasound on the detector strain sensitivity is discussed in \cref{subsubsec:newtonian_atmospheric}.

Other mechanisms of atmospheric noise generation include spatially varying temperature fields that move near the test mass due to wind, and pressure fluctuations generated by turbulent mixing (the aeroacoustic effect), but these noise sources are unlikely to be significant above \SI{5}{\Hz}~\cite{2019LRR....22....6H}.
Finally, details of the dimension and shape of the buildings housing the test masses can alter the above noise sources (e.g., by excluding large density fluctuations close to the test masses), but have the potential to introduce extra noise due to local vortices~\cite{2008CQGra..25l5011C}.
We do not consider details of the test mass buildings here, but note that proper design will be needed to ensure that atmospherically induced noise is kept to a minimum.
Accurately modeling the Newtonian noise contribution below \SI{5}{\Hz} is an area of ongoing research, and we do not attempt a detailed noise analysis in this frequency band.

\subsection{Vacuum system}
\label{subsec:facility_tube}

The design of the Cosmic Explorer vacuum system, including the beam tube infrastructure and test mass chambers, has not been determined. However, in \cref{subsec:scatter} we determine that a beam tube diameter of \SI{120}{\cm} with a similar acceleration spectrum as the LIGO beam tube motion is likely sufficient to keep noise from back-scattered light well below the total Cosmic Explorer noise, though this will be reevaluated once forward-scattering effects are accounted for.

Although the beam tubes and test mass chambers are evacuated, the small amount of residual gas causes noise in the detector through two mechanisms discussed in \cref{subsec:resgas}. The first is optical path length fluctuation due to the polarizability of the molecules in the beam tubes passing through the laser beam~\cite{1996magr.meet.1434Z,2002JVST...20.1237T}, and the second is test mass motion due to momentum transfer from the gas molecules in the chambers~\cite{2010PhLA..374.3365C,2011PhRvD..84f3007D}. Achieving low pressures is more challenging in the chambers than in the beam tubes because the chambers will be periodically opened in order to make modifications to the detector. We thus set the vacuum system requirements such that the total residual gas noise for the \SI{1}{\um} technology is a factor of three below the CE2 design sensitivity at \SI{10}{\Hz} and a factor of five below the design sensitivity at \SI{100}{\Hz}.

In this work we assume that the total vacuum pressure in both the tubes and chambers is dominated by molecular hydrogen, water, molecular nitrogen, and molecular oxygen with each species contributing equally to the total gas noise. Under these assumptions, the above noise requirements translate into requirements on the partial pressures in the beam tubes of
$P_\mr{H_2}=\SI[round-mode=figures]{44.0}{\nano\pascal}$,
$P_\mr{H_2O}=\SI[round-mode=figures]{4.0}{\nano\pascal}$,
$P_\mr{N_2}=\SI[round-mode=figures]{2.5}{\nano\pascal}$,
and
$P_\mr{O_2}=\SI[round-mode=figures]{2.8}{\nano\pascal}$,
for hydrogen, water, nitrogen, and oxygen, respectively, and on the partial pressures in the test mass chambers of
$P_\mr{H_2}=\SI[round-mode=figures]{410.0}{\nano\pascal}$,
$P_\mr{H_2O}=\SI[round-mode=figures]{140.0}{\nano\pascal}$,
$P_\mr{N_2}=\SI[round-mode=figures]{110.0}{\nano\pascal}$,
and
$P_\mr{O_2}=\SI[round-mode=figures]{100.0}{\nano\pascal}$. It is also important that the hydrocarbons are kept low enough that they do not contaminate the mirror surface and cause excess optical loss.

\section{Noise estimates}
\label{sec:noise}

This section describes noise terms that contribute to the limit of the low-frequency performance of Cosmic Explorer.

\subsection{Suspension thermal noise}
\label{subsec:susthermal}

\begin{figure}
  \centering
  \includegraphics[width=\figwidth]{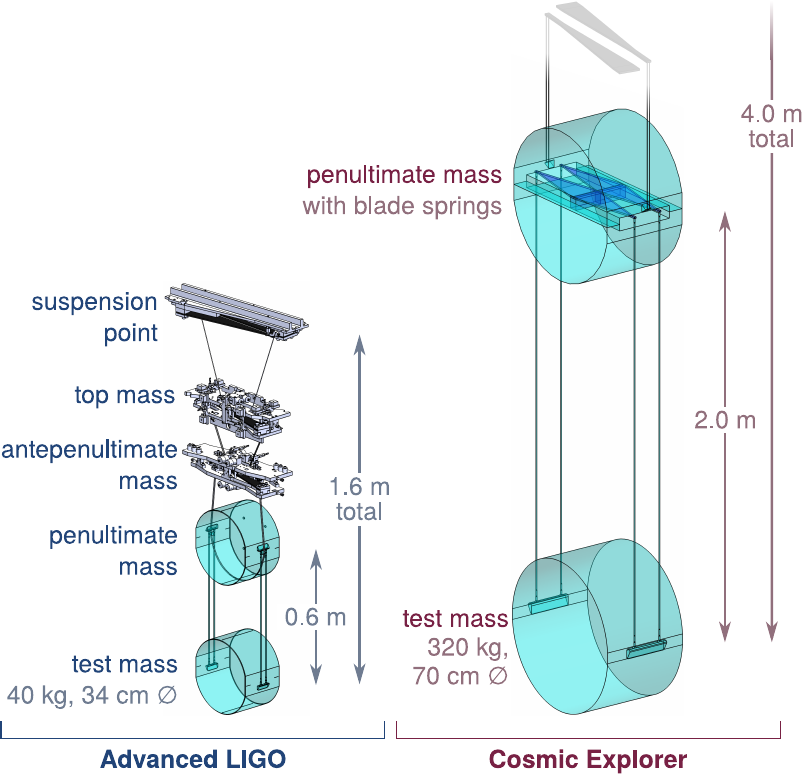}
  \caption{\textit{Left:} schematic of the Advanced LIGO quadruple suspensions. \textit{Right:} one design concept for the final two stages of a Cosmic Explorer silica suspension for a \SI{70}{\cm} diameter fused silica test mass. The components shown in blue are fused silica. In particular, the test masses, PUMs, and fibers between the two are are fused silica as are the blade springs on the CE PUM. The components shown in black are maraging steel blade springs. The components shown in silver are the other steel components on the LIGO suspensions. The silicon CE suspensions have silicon ribbons, silicon blade springs on the PUM, and a \SI{80}{\cm} diameter test mass. Note that only the final two stages of the CE suspensions are shown; the full suspension would be similar to LIGO's but would have \SI{4}{\m} total length rather than \SI{1.65}{\m}.}
  \label{fig:suspensions}
\end{figure}

\begin{figure*}
  \centering
   \includegraphics[width=\figwidth]{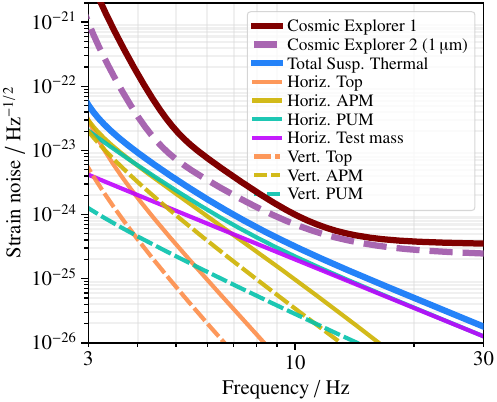}
  \includegraphics[width=\figwidth]{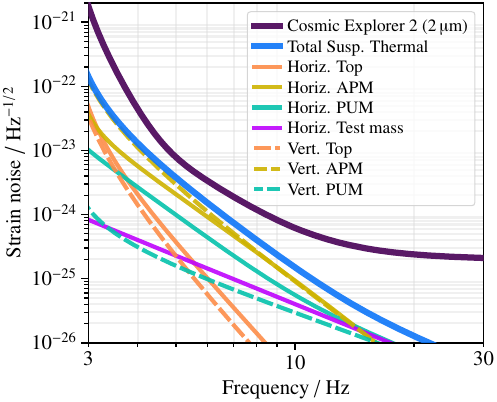}
  \caption{Contribution of each stage of the test mass quadruple suspension to the total suspension thermal noise for the \SI{1}{\um} technology (left), which would be common for CE1 and CE2, and for the \SI{2}{\um} technology (right).}
  \label{fig:suspension_thermal}
\end{figure*}

The baseline Cosmic Explorer design assumes scaled-up versions of the quadruple pendulum
suspensions used in LIGO~\cite{2012CQGra..29w5004A} and planned for
Voyager~\cite{2020arXiv200111173A}, along with a few modifications, to decrease
the seismic and suspension thermal noises. The left panel of \cref{fig:suspensions} shows a diagram of the LIGO suspensions. Suspension thermal noise is related
to the mechanical response of the suspensions through the fluctuation
dissipation
theorem~\cite{1951PhRv...83...34C,1952PhRv...86..702C,1994ASAJ...96..207G,2000CQGra..17.4409G}
as $S(f)\propto T\im \chi / f$, where $\chi$ is the mechanical susceptibility.

In order to minimize thermal noise, the final suspension stage---consisting of
the penultimate mass (PUM), the test mass, and the fibers or ribbons between
them---is monolithic; for the \SI{1}{\um} technology, the material is
room-temperature fused silica, and for \SI{2}{\um} technology, the material is
cryogenic silicon. The top two masses, called the top mass and the
antepenultimate mass (APM), are room-temperature maraging steel for both
wavelengths. In order to lower the vertical suspension resonances, the top
three stages are suspended by steel wires from steel blade springs attached to
the stage above.

In order to further reduce the resonances, the test masses are suspended by a
final set of blade springs attached to the PUM made from the same material as
the PUM and test mass. One concept for the design of this final stage is shown in the right panel of \cref{fig:suspensions}.
The stress and spring constant of the blade can be calculated with beam
theory~\cite{Gere-Timoshenko1997} by approximating it as a rectangular
cantilever of length $\ell$, width $w$, and thickness $h$. The maximum stress
$\sigma_\text{max}\propto \ell/wh^2$ occurs at the clamp, and the spring constant
$k\propto wh^3/\ell^3$ is the ratio of the load suspended by the blade to its
maximum deflection at the tip. The blade dimensions should be chosen to
minimize $k$ while keeping the maximum stress below a safety factor of the
breaking stress of the blade.

For the \SI{1}{\um} technology, as with LIGO, the silica test mass is suspended from the PUM by four silica fibers welded to the test mass~\cite{2012CQGra..29w5004A}; in Cosmic Explorer they are welded at the top to the blade springs while in LIGO they are welded
directly to the PUM. The contribution of the loss angle $\phi$ to the imaginary
part of the horizontal spring constant $\im k\propto \phi/D$ is reduced by the
dilution factor $D\propto I^{-1/2}$, where $I$ is the cross-sectional area
moment of inertia of the fiber or ribbon~\cite{1990PhRvD..42.2437S,
2000PhLA..270..108G, 2000CQGra..17.4409G}. Since $I\propto r^4$ for a fiber of radius
$r$, it is advantageous to make the radius as small as the breaking stress of
the fiber allows. Maximizing the stress in the fiber in this way has the added
benefit of reducing the contribution of the fiber to the vertical spring
constant and increasing the frequency of the first violin mode, which is
proportional to $\sigma^{1/2}$.

The thermoelastic noise of the fiber has two contributions: one from thermal
expansion and one from the temperature dependence of the Young modulus. These
two contributions cancel when the fiber stress is appropriately chosen. Thus, a
tapered fiber is used with a larger radius at the ends (where the most bending,
and therefore the most loss, occurs) chosen to give the stress necessary to
cancel the thermoelastic noise, and a smaller radius along the length of the
fiber chosen to maximize the stress~\cite{2012CQGra..29w5004A}.

For the \SI{2}{\um} technology, as with Voyager, the silicon test mass is suspended by four silicon ribbons welded to the test mass at the bottom and to the blade springs at the top. Since the ribbons are held near the zero-crossing of the thermal expansion
coefficient, the thermoelastic noise in the ribbons cannot be canceled by
choice of stress as is done for the fused silica fibers. The ribbon dimensions
are therefore chosen to maximize the stress along the entire length of the
ribbon. Since $I\propto wh^3$ for a ribbon of width $w$ and thickness $h$, a
width-to-thickness ratio of 10:1 is chosen to soften the pendulum in the
horizontal direction and to increase the gravitational dilution.

The suspension design also determines the seismic noise, discussed below in
\cref{subsec:seismic}, since the suspensions provide passive $1/f^8$ filtering
of seismic noise above all of the longitudinal, vertical, and angular
resonances. To reduce both seismic and suspension thermal noise, it is thus
advantageous to make the suspensions as soft as possible and to lower their
resonances.

To achieve this goal, the total allowable height of the suspensions for all
technologies has been increased to \SI{4}{\m} and the total mass increased to
\SI{1500}{\kg}. Within these constraints, in an analysis similar to that done
for Voyager~\cite{2020arXiv200111173A}, the lengths and masses of the silica
and silicon suspension stages have been optimized to minimize the sum of these
noises over the frequency band of \SIrange[range-units=single]{4}{15}{\Hz}.

\cref{fig:suspension_thermal} shows the contributions of each stage to
the total suspension thermal noise. The silica suspensions are dominated
by the horizontal noise of the PUM and test mass above about \SI{10}{\Hz} with contributions from the horizontal noise of the APM below. The silicon suspensions
are dominated by vertical noise of the APM below about \SI{7}{\Hz}, above
which the horizontal and vertical noises of the PUM and test mass also start to
become important. The addition of blade springs lowers the first vertical mode
thus reducing the vertical thermal noises, most importantly from the APM.

The maximum stress that the blade springs, fibers, and ribbons can tolerate is
an important material property in the design of the suspensions, and it is
difficult to predict what will be possible on a timescale of decades. The maximum
stress of the LIGO silica fibers is
\SI{800}{\mega\pascal}~\cite{2012CQGra..29w5004A}, which provides a safety factor
of about six for the breaking stress of fibers realized at the time the LIGO
suspensions were designed~\cite{2012JNCS..358.1699T}. Recent improvements to fused
silica fiber fabrication suggest that fibers can be made with stresses of
\SI{1.2}{\giga\pascal}, which provides a safety factor of about
three~\cite{2019CQGra..36r5018L}. The Cosmic Explorer fused silica suspensions use
this \SI{1.2}{\giga\pascal} for the fibers and tentatively set the maximum
blade spring stress to be \SI{800}{\mega\pascal}.

The silicon studies most relevant to the suspensions discussed in this section
find that the tensile strength depends on the surface treatment and edge
quality, with average breaking stresses measured ranging from
\SIrange[range-units=single]{100}{400}{\mega\pascal} and individual samples
observed as high as \SI{700}{\mega\pascal}~\cite{2019CQGra..36r5018L,
2014CQGra..31b5017C}. Cosmic Explorer tentatively sets a maximum stress of
\SI{400}{\mega\pascal} for both the blades and ribbons while Voyager uses a
more conservative \SI{100}{\mega\pascal}~\cite{2020arXiv200111173A}.
Nevertheless, larger stresses have been observed in other contexts. Stresses of
\SIrange[range-units=single]{3}{5}{\giga\pascal} have been observed in silicon
wafers~\cite{1987JCrGr..85...83M} and micro-scale MEMS devices have realized
fracture stresses in excess of \SI{1}{\giga\pascal} and stresses of up to
\SI{10}{\giga\pascal} have been realized in nano-scale
devices~\cite{2015ApPRv...2b1303D}.

No blade springs have yet been constructed out of either silica or silicon.
Developing this technology and techniques for manufacturing highly stressed
materials is a critical area of research and development in realizing the
low-frequency sensitivity of Cosmic Explorer. Alternatives to blades springs,
such as geometric antisprings~\cite{2002NIMPA.487..652C}, should also be
developed in parallel. Additionally, no experiment on earth has ever directly
measured (low) suspension thermal noise.

\subsection{Seismic noise}
\label{subsec:seismic}

\begin{figure}
    \centering
    \includegraphics[width=\figwidth]{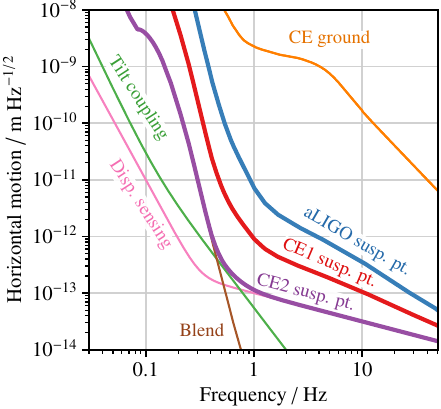}
    \caption{Horizontal motion of the Cosmic Explorer suspension point, shown for both CE1 and CE2.
    CE1 assumes seismic isolation that is moderately improved compared to Advanced LIGO.
    CE2 assumes further improvements to the seismic isolation using novel inertial sensing technology~\cite{2019CQGra..36x5006M}.
    A simplified budget of the CE2 motion is also shown, along with the CE ground motion model (\cref{fig:ground_spectra}).}
    \label{fig:seismic_isolation}
\end{figure}

Like Advanced LIGO~\cite{2015CQGra..32r5003M,2012CQGra..29w5004A}, Cosmic Explorer will suppress seismic noise with passive and active techniques.
The suspensions described in \cref{subsec:susthermal} passively filter the seismic noise with a $1/f^8$ slope in amplitude above the suspension resonances, which have been reduced with the optimization described there. Even so, in order to achieve the required seismic noise suppression, the motion of the optical table supporting the suspension will be actively suppressed with a combination of inertial sensors and position sensors. The seismic isolation of the Cosmic Explorer~1 and~2 suspension point is shown in \cref{fig:seismic_isolation}.

For CE1, we assume an isolation performance that is moderately improved compared to Advanced LIGO~\cite{2015CQGra..32r5003M}.
At ${\sim}\SI{10}{\Hz}$ we assume a threefold improvement, and at ${\sim}\SI{1}{\Hz}$ a tenfold improvement, though to directly increase the seismic isolation the improvement is only needed down to \SI{5}{\Hz}; seismic isolation improvements below the gravitational-wave band will, however, lessen the requirements on the interferometer control system.
The improvement could come, for example, by combining the mechanics of a conventional geophone (GS13) with an interferometric proof mass readout~\cite{2018CQGra..35i5007C}.
The noise below \SI{1}{\Hz} is residual ground motion that comes from the inclusion of a position sensor signal to lock the suspension point to the ground on long timescales (also referred to as ``blending'').
Additionally, the horizontal inertial sensing is susceptible to contamination from ground tilt, and should therefore be paired with low-noise tiltmeters~\cite{2014RScI...85a5005V}.
This is motivated by studies at LIGO Hanford that have shown significant tilt-to-interferometer strain coupling after active seismic isolation~\cite{2018PhRvL.121v1104C}.\footnote{Lowering the tilt coupling, along with mitigating gravity gradient fluctuations from the atmosphere, is an important motivator for carefully designed buildings~\cite{2020CQGra..37r5018R}.}

For CE2, we assume that improvements in inertial sensing will yield another threefold noise improvement at \SI{10}{\Hz} and a tenfold improvement at \SI{1}{\Hz}, again with the improvement only needed down to \SI{5}{\Hz} to achieve a direct seismic isolation improvement.
A variety of designs have been proposed, but common themes include a monolithic proof mass assembly to reduce thermal noise and an optical displacement sensor to reduce readout noise.
van~Heijningen et~al.\ demonstrated a monolithic accelerometer combined with an interferometric readout that reached a noise floor of \SI{8e-15}{\m\,\Hz^{-1/2}} above \SI{30}{\Hz}; this should reach \SI{e-15}{\m\,\Hz^{-1/2}} above \SI{10}{\Hz} with continued development~\cite{vanHeijningen2018}.
A proposed superconducting niobium upgrade to this system would reduce eddy current damping and greatly improve suspension thermal noise allowing, in principle, \SI{e-15}{\m\,\Hz^{-1/2}} above \SI{1}{\Hz}~\cite{2020JInst..15P6034V}.
However, such a device has yet to be demonstrated, and would operate at temperatures below \SI{9.2}{\K}, requiring additional cooling with respect to the Cosmic Explorer cryogenic environment, and would require a low-noise tiltmeter.
Development of novel six-dimensional inertial isolators with optical readouts is also progressing~\cite{2019CQGra..36x5006M}, and their use with the existing LIGO facilities and Advanced LIGO isolation infrastructure has been explored~\cite{2018PhRvL.120n1102Y}.
These sensors would provide the additional benefit of sensing tilt.
Additionally, the improved low-frequency noise of the inertial sensors leads to less reliance on the low-frequency position sensor signal, thereby lessening the contamination from residual ground motion.
 
\subsection{Newtonian noise}
\label{subsec:newtonian}

\begin{figure*}
    \centering
    \includegraphics[width=\figwidth]{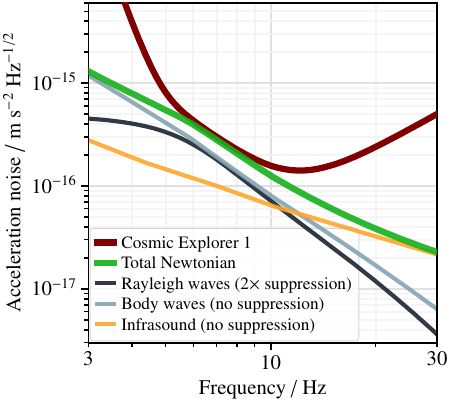}
    \includegraphics[width=\figwidth]{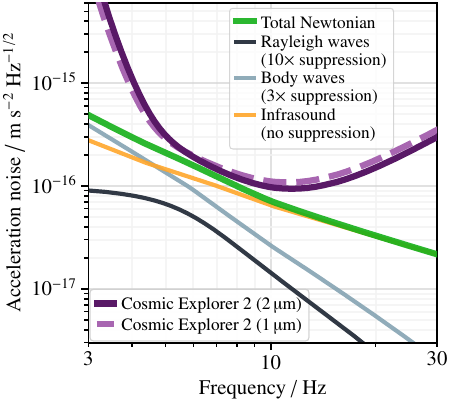}
    \caption{
    Newtonian noise estimates for Cosmic Explorer.  For CE1, the Rayleigh wave
content is assumed to be suppressed by a factor of 2 in amplitude below the
ground motion shown in \cref{fig:ground_spectra}, either through offline
subtraction or local mitigation (e.g., excavation as described in
\cref{subsubsec:newtonian_mitigation}) in the immediate vicinity of the test
mass.  The P- and S-wave amplitudes are each assumed to be a factor of 10
higher than the Peterson low-noise model~\cite{Peterson1993}.  For CE2, the
Rayleigh wave content is assumed to be suppressed by a factor of 10 in
amplitude, and the body wave content is suppressed by a factor of 3 in
amplitude.  The infrasound amplitude is taken from the Bowman
model~\cite{2005GeoRL..32.9803B}.
    }
    \label{fig:newtonian}
\end{figure*}

\begin{figure}
    \centering
    \includegraphics[width=\figwidth]{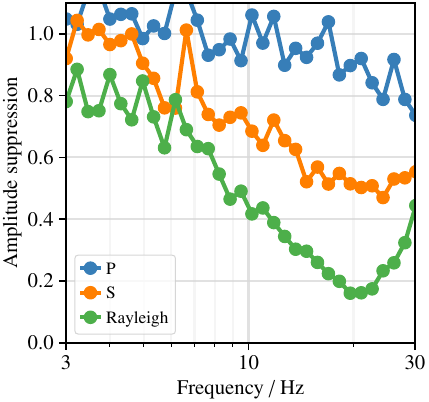}
    \caption{
    Seismic Newtonian-noise reduction amplitudes for P, S, and Rayleigh waves
achieved by removing ground from underneath the test mass to make a
\SI{11}{\meter} wide and \SI{4}{\meter} deep recess.  This reduction estimate
is computed using the Born approximation, which may affect the validity of the
Rayleigh-wave reduction estimate above \SI{15}{\Hz}~\cite{2014CQGra..31r5011H};
the body-wave reduction estimate should not be significantly affected.  The
scatter in the curves is due to the finite number of waves simulated and the
finite size of the numerical grid.
    }
    \label{fig:NNholes}
\end{figure}

Previous studies of Newtonian noise for Cosmic Explorer considered only the
contribution from seismic Rayleigh waves, and assumed a Rayleigh-wave noise
amplitude equal to that of the existing LIGO
facilities~\cite{2017CQGra..34d4001A}.  Here we refine that estimate and
additionally include the contributions from seismic body waves and from
atmospheric infrasound.  We start with analytical formulae available in the
literature for the infinite half-space, and then additionally we consider
numerical simulations that account for trenches that can reduce Newtonian noise
relative to the half-space solutions.  The Newtonian noise estimates are shown
in \cref{fig:newtonian}.

\subsubsection{Seismic Newtonian noise}
\label{subsubsec:newtonian_seismic}

As described in Sec.~\ref{subsec:facility_seismic}, we assume that compared to
LIGO, the Cosmic Explorer facility will have a lower Rayleigh-wave noise in the
anthropogenic band: $\SI{1}{\um\,\second^{-2}\,\Hz^{-1/2}}$ above \SI{5}{\Hz}.
We also assume a body-wave noise amplitude equal to
$\SI{0.3}{\um\,\second^{-2}\,\Hz^{-1/2}}$ above \SI{5}{\Hz}, equipartitioned
among P-waves, vertically polarized S-waves, and horizontally polarized S-waves.

To compute the Newtonian noise from seismic and infrasonic density
fluctuations, we employ the formulae from Harms~\cite{2019LRR....22....6H},
which are valid for a test mass suspended above a homogeneous, isotropic
elastic half-space.  We therefore do not consider effect of
stratigraphy, other ground anisotropies, the interaction with structures, or
the interconversion of different types of seismic waves.  These features will
need to be accounted for to get a full understanding of the behavior of the
local seismic field and hence the Newtonian noise level.  For
CE1, we have assumed that the effect of seismic Newtonian noise can be
mitigated (\cref{subsubsec:newtonian_mitigation}) with
$\num{2}{\times}$ amplitude suppression of
Rayleigh waves.  The result in \cref{fig:newtonian} shows that CE1 is limited
by seismic Newtonian noise from \num{5}--\SI{10}{\Hz}, with a secondary
contribution from infrasound.  For CE2, we have assumed that seismic Newtonian
noise can be further mitigated with
$\num{10}{\times}$ amplitude suppression for
Rayleigh waves and $3\times$ amplitude suppression for body waves; the result
in \cref{fig:newtonian} shows that CE2 is then limited by atmospheric Newtonian
noise, described below.

\subsubsection{Atmospheric Newtonian noise}
\label{subsubsec:newtonian_atmospheric}

As mentioned in \cref{subsec:facility_atmospheric}, we assume the Cosmic
Explorer facility has a typical infrasound spectrum of
$\SI{1}{\milli\pascal\,\Hz^{-1/2}}$; this is an extrapolation from long-term
global infrasound data, available below
\SI{10}{\Hz}~\cite{2005GeoRL..32.9803B}, and assumes no significant
contribution from site infrastructure.

To compute the Newtonian noise induced by infrasound fluctuations, we use the
calculation in Harms~\cite{2019LRR....22....6H}, which is valid for a test mass
immersed in a fluid half-space.  The result is shown in \cref{fig:newtonian}.
For both stages of Cosmic Explorer, no suppression is assumed.

As mentioned in \cref{subsec:facility_atmospheric}, we do not include other
processes besides infrasound that produce density fluctuations in the
atmosphere, such as advected temperature fluctuations or aeroacoustic noise,
because we expect the Newtonian noise induced by these processes to be
negligible above a few hertz.

\subsubsection{Mitigation strategies}
\label{subsubsec:newtonian_mitigation}

Unlike mechanically coupled seismic and acoustic noise, which can be strongly attenuated
by suspending and inertially isolating the test mass inside a vacuum chamber,
the Newtonian effect of seismic and acoustic fluctuations cannot be attenuated
except by reducing the fluctuation amplitude, increasing the distance from the
fluctuations to the test mass, or using auxiliary sensors to estimate the
Newtonian contribution to the detector strain channel.  Newtonian noise
mitigation therefore requires a different set of techniques than for mechanical
isolation, and the amount of achievable suppression will not be as great.

CE1 calls for mitigating the seismic Rayleigh-wave Newtonian noise by a factor
of \num{2} in amplitude; CE2 calls for mitigating
the seismic Rayleigh-wave Newtonian noise by a factor of
\num{10} in amplitude, and the seismic body-wave
Newtonian noise by a factor of 3 in amplitude.  This mitigation could be
achieved by several means, potentially used in concert:

\begin{enumerate}

    \item \emph{Seismometer array subtraction.} Arrays of seismometers can be
used to estimate the seismic field in the vicinity of the test mass and thereby
subtract Newtonian noise from the gravitational-wave
channel~\cite{2012PhRvD..86j2001D}.  A proof-of-principle experiment to
subtract ground motion from a tiltmeter signal achieved a tenfold suppression
in the region
\SIrange[range-phrase=--]{10}{20}{\Hz}~\cite{2018PhRvL.121v1104C}.

    \item \emph{Excavation underneath the test masses.}
\label{item:NNexcavation} Nearby density and displacement fluctuations can be
suppressed simply by removing earth from the vicinity the test mass, replacing
it with a lightweight fill material such as extruded polystyrene if necessary.
Harms and Hild~\cite{2014CQGra..31r5011H} computed the suppression of
Rayleigh-wave Newtonian noise from a \SI{11}{\meter} wide and \SI{4}{\meter}
deep hemispherical recess, and here we repeat their analysis to additionally
include the effect of the recess on P- and S-waves.  The result is shown in
\cref{fig:NNholes}, showing that moderate reduction of Rayleigh waves can be
achieved near and above \SI{10}{\Hz}, while the reduction of body waves is less
significant.

    \item \emph{Topography and seismic metamaterials.}
\label{item:NNmetamaterials} Seismic metamaterials could be built
to deflect or dissipate seismic waves before they arrive at the test mass,
potentially suppressing surface wave amplitudes by a factor of a
few~\cite{2014PhRvL.112m3901B,2016NatSR...627717C,2016NatSR...639356P,roux2018toward,2019APS..APRR11006K,2020PhRvP..13c4055Z}.
Similarly, berms, ditches, and other nearby topographic features can affect the
propagation of seismic waves, and thus the Newtonian noise level.

\end{enumerate}

No mitigation of infrasound noise is assumed, and thus infrasound is considered
a sensitivity limit of the Cosmic Explorer facility.  Tropospheric LIDAR, which
would otherwise be well-suited to three-dimensional estimation of atmospheric
fluctuations, would require sensitivity improvements of several orders of
magnitude in order to sense and subtract infrasound~\cite{FiorucciGWADW2019}.
Baffling or otherwise acoustically isolating the interior of the test mass
building may be able to reduce the infrasound Newtonian noise below the outdoor
value at a discrete set of frequencies~\cite{2018PhRvD..97f2003F}.  A true
cutoff for infrasound noise could be engineered by burying the test mass a
depth $d$ below ground, which would suppress the noise by $\rme^{-d f /
c_\text{s}}$, where $c_\text{s}$ is the speed of sound; however, to achieve
significant suppression for $f \ge \SI{5}{\Hz}$ would require $d \ge
\SI{65}{\meter}$.  Additionally, underground operation requires a reassessment
of the Newtonian noise, since the detector would operate in the bulk of the
ground rather than on the surface.
 
\subsection{Test mass thermal noise}
\label{subsec:TMthermal}

\begin{figure*}
	\includegraphics[width=\figwidth]{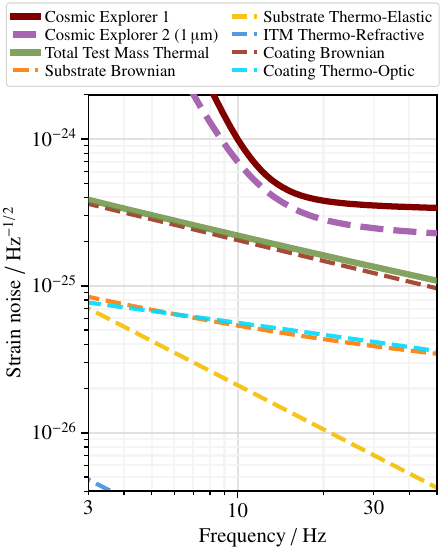}
	\includegraphics[width=\figwidth]{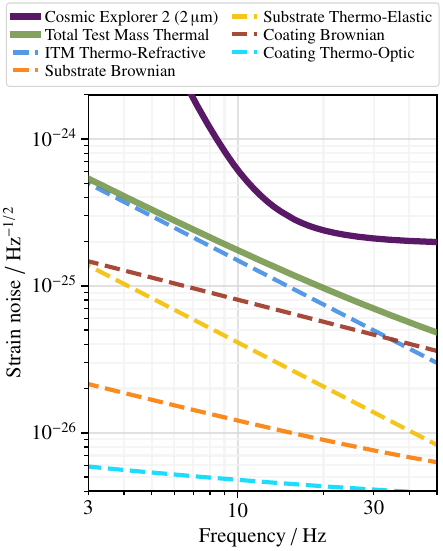}
\caption{Thermal noise levels, and individual noise contributions to them, of the test mass substrates and coatings for the \SI{1}{\um} technology (left), which would be common for CE1 and CE2, and for the \SI{2}{\um} technology (right).} \label{fig:substrate_thermal}
\end{figure*}

Cosmic Explorer will use heavy, high-quality test masses which are turned into
high-reflectivity Bragg mirrors by coating the test mass surface with multiple
layers of dielectric films. By alternating between
high- and low-refractive-index materials, and depositing the layers to a
thickness that is the same scale as the laser wavelength, the coating creates
the conditions for repeated thin-film interference of the laser
beam~\cite{1968imos.book.....F}. The performance of the coating depends on the
optical, mechanical, and thermal properties of the materials, which therefore
must be chosen with some care~\cite{2012octn.book.....H}.
 
The \SI{1}{\um} coating technology will mostly be that of LIGO A+:
room-temperature fused silica substrates and coating technology being developed
for A+~\cite{2020ApOpt..59A.229G}.  Current research aimed at improving the
thermal noise of room-temperature coatings holds promise to result in improved
coatings for A+ and thus the \SI{1}{\um} CE
technology~\cite{vajente2020amorphous}.  The \SI{2}{\um} technology will mostly
be that of LIGO Voyager: crystalline silicon substrates operated at
\SI{123}{\kelvin}, with coating materials that offer improved thermal noise
performance over the \SI{1}{\um} technology. 

Estimated thermal noises associated with the Cosmic Explorer test masses and
their coatings are shown in \cref{fig:substrate_thermal} and the individual
noises are discussed below. Neither the A+ nor the Voyager coating designs
have been finalized, so in this work we have made assumptions about the high-
and low-index material pairs. Depending on the progression of coating research
in the next decade, it is possible that the coatings for CE1 or CE2 may be
different from what is presented here, and could potentially use three or more
materials to provide more flexibility to simultaneously optimize the optical
and thermal noise properties of the
mirrors~\cite{2015PhRvD..91d2002Y,2020PhRvL.125a1102T}.

\subsubsection{Substrates}
\label{subsubsec:TMthermal_substrate}

Cosmic Explorer will use \SI{320}{\kg} test mass substrates; this
comes from the desire to make quantum radiation-pressure noise subdominant to
other noise sources and the necessity of having large test masses to
accommodate the large diameter beams of a nearly diffraction limited
\SI{40}{\km} long arm cavity.
There are several sources of thermal noise in test mass substrates: mechanical
(Brownian) noise, thermoelastic noise, and thermorefractive noise.

Brownian fluctuation causes a displacement of the mirror surface with a power
spectrum $S(f) \propto T \phi / w f$, where $T$ is the test mass temperature,
$w$ is the spot size of the beam, and $\phi$ is the mechanical loss of the
substrate material; there are order unity corrections due to the finite size of
the test mass and additional loss on the test mass
surface~\cite{2018PhRvD..98l2001G}.

\begin{figure*}
	\includegraphics[width=\figwidth]{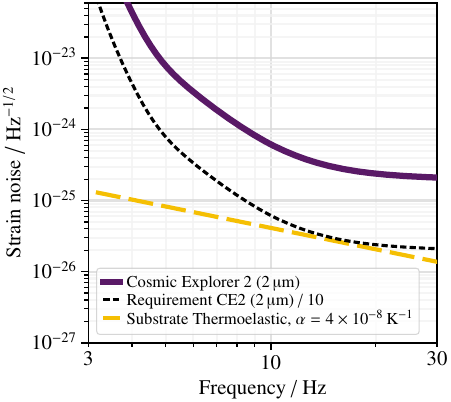}
	\includegraphics[width=\figwidth]{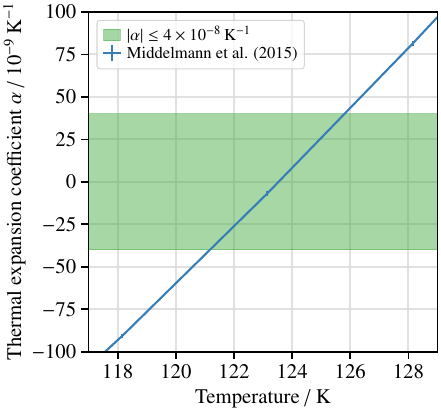}
    \caption{
    Left: Amplitude spectral sensitivity of CE2 realized by the cryogenic silicon
    \SI{2}{\um} technology compared with the estimated thermoelastic noise of the
    silicon test mass substrates for $\alpha=\SI{4e-8}{\per\kelvin}$.  The
    requirement that thermoelastic noise be a factor of ten below the CE2 design
    curve is met when $\alpha=\pm\SI{4e-8}{\per\kelvin}$. Right: Coefficient of
    thermal expansion of crystalline silicon versus temperature measured by
    Middelmann et al.~\cite{2015PhRvB..92q4113M}, zoomed to show the data points
    and error bars around the zero crossing at 123.5\,K. The green region indicates
    the required $|\alpha|\le\SI{4e-8}{\per\kelvin}$, corresponding to a
    temperature accuracy of about $\pm \SI{2.3}{\per\kelvin}$.
}
    \label{fig:cte}
\end{figure*}

Thermoelastic noise is driven by thermodynamic fluctuations that cause
displacement of the test masses via the coefficient of thermal expansion,
$\alpha$~\cite{1999PhLA..264....1B}: the spectrum of the test mass surface
displacement due to these fluctuations is $S(f) \propto T^2 \kappa \alpha^2 /
w^3 f^2$, where $\kappa$ is the thermal conductivity of the substrate.  For
fused silica, the contribution of substrate thermoelastic noise to the total
instrument noise is negligible.  In order to prevent the substrate
thermoelastic noise of silicon from making a significant contribution, the
substrate temperature must be controlled to near the zero-crossing of the
thermal-expansion coefficient~\cite{2020arXiv200111173A,voyager-cryo}.  The
left panel of \cref{fig:cte} shows that $|\alpha| \le \SI{4e-8}{\K^{-1}}$ meets
the requirement for thermoelastic noise to be an order of magnitude below the
total design sensitivity.
Based on models~\cite{2018PNAS..115.1992K} and
measurements~\cite{2015PhRvB..92q4113M} of the temperature dependence of
$\alpha$, this constraint on $\alpha$ translates to a temperature control
requirement of $\pm \SI{2.3}{\K}$ relative to the zero-crossing temperature of
$\alpha$.
This temperature control accuracy is also sufficient to keep
thermoelastic noise of the silicon components of the suspension from
contributing significantly to the total low frequency suspension thermal noise
for the \SI{2}{\um} technology as shown in \cref{fig:suspension_thermal}.

\begin{table*}
    \centering
    \begin{tabular}{
        r
        r
        l
        S[table-align-text-post=false]S[table-align-text-post=false,
]l}
        \toprule
        &
            \multicolumn{2}{c}{\textbf{Quantity}}   &
            {\textbf{\SI{1}{\um} Technology}}    &
            {\textbf{\SI{2}{\um} Technology}}    &
            \textbf{Remarks}    \\
        \hline
        Substrate   &
            Material    &
            &
            {Fused silica}    &
            {Crystalline silicon} &
            \\
        & 
            Temperature &
            $T$ &
            293\,\si{\kelvin}   &
            123\,\si{\kelvin}   &
            To within $\pm\SI{2.3}{\kelvin}$ for CE2 \\
        &
            Diameter &
            &
            \sisetup{round-mode=figures,round-precision=2}70.0\,\si{\cm}   &
            \sisetup{round-mode=figures,round-precision=2}80.0\,\si{\cm}   &
            \\
        &
            Thickness &
            $H$ &
            \sisetup{round-mode=figures,round-precision=2}37.8\,\si{\cm}   &
            \sisetup{round-mode=figures,round-precision=2}27.3\,\si{\cm}   &
            \\
        &
            Mass &
            $M$ &
            320\,\si{\kg}   &
            320\,\si{\kg}   &
            \\
        &
            Thermal expansion coeff. &
            $\alpha$ &
            0.39\,\si{ppm\ \kelvin^{-1}}   &
            0.04\,\si{ppm\ \kelvin^{-1}}   &
            See remark on $T$ \\
        &
            Refractive index &
            $n$ &
            1.45   &
            3.5   &
            \\
        &
            Thermorefractive coeff. &
            $\beta$ &
            9.6\,\si{ppm\ \kelvin^{-1}}   &
            \sisetup{round-mode=figures}100.0\,\si{ppm\ \kelvin^{-1}}   &
            \\
        &
            Thermal conductivity &
            $\kappa$ &
            1.38\,\si{\watt\ \meter^{-1}\,\kelvin^{-1}}   &
            700\,\si{\watt\ \meter^{-1}\,\kelvin^{-1}}   &
            \\
        \hline
        Coating &
            Materials  &
            &
            {SiO$_2$ / TBD} &
            {SiO$_2$ / aSi} &
            Low index / high index \\
        &
            Refractive indices  &
            &
            {1.45   /
            2.07}  &
            {1.44   /
            3.5}  &
            \\
        &
            Loss angles  &
            &
            {\num{2.3e-05}   /
            \num{7e-05}}  &
            {\sisetup{scientific-notation=true}\num{0.0001}   /
            \num{3e-05}}  &
            \\
        &
            ITM coating layers  &
            &
            16  &
            11  &
            \\
        &
            ETM coating layers  &
            &
            38  &
            15  &
            \\
        \hline
        Optical &
            Vacuum wavelength   &
                $\lambda$       &
                \sisetup{round-mode=figures,round-precision=1}1.0\,\si{\um} &
                \sisetup{round-mode=figures,round-precision=1}2.0\,\si{\um} &
                \\
        &
            ITM spot size       &
                $w_\text{i}$    &
                \sisetup{round-mode=figures,round-precision=2}12.0\,\si{\cm} &
                \sisetup{round-mode=figures,round-precision=2}16.4\,\si{\cm} &
                $1/\rme^2$ intensity radius \\
        &
            ETM spot size       &
                $w_\text{e}$    &
                \sisetup{round-mode=figures,round-precision=2}12.0\,\si{\cm} &
                \sisetup{round-mode=figures,round-precision=2}16.4\,\si{\cm} &
                \\
        &
            ITM transmissivity  &
                $\mathcal{T}_\text{i}$    &
                1.4\,\si{\percent} &
                1.4\,\si{\percent} &
                \\
        &
            ETM transmissivity  &
                $\mathcal{T}_\text{e}$    &
                \sisetup{round-mode=figures,round-precision=1}5.0\,\si{ppm} &
                \sisetup{round-mode=figures,round-precision=1}5.0\,\si{ppm} &
                \\
    \botrule
    \end{tabular}
    \caption{
    Requirements for the substrate, coating, and optical properties of the Cosmic Explorer test masses. The high-index coating material for the room-temperature technology is not known, so it has been assumed to have the same properties as the titania-doped tantala used in current detectors, but with a mechanical loss such that the overall coating loss is four times lower than the current Advanced LIGO coating loss.
    }
    \label{tab:tm_params}
\end{table*}

To achieve $\pm\SI{2.3}{\K}$ temperature control, it may be sufficient to
control the test mass temperatures to a fixed value (for example using the
frequency of the internal modes of the silicon test masses as a reference for
temperature), or it may be necessary to determine the set temperature based on
minimizing the observed noise or by actively measuring the substrates' $\alpha$
values.  The sign change of $\alpha$ around the zero crossing allows for a
signed error-signal that would enable negative feedback control.  Typical room
temperature variations achieved at the current LIGO observatories are of order
$\pm\SI{1}{\kelvin}$, and even better accuracy should be achievable with
feedback control~\cite{surf-cryo,voyager-cryo}.  Temperature gradients due to
heating from the environment and from absorbed laser power also need to be
considered.  If a power $P_\text{abs}$ is absorbed on some area $A$ of the test
mass and dissipates into the substrate, the resulting temperature variation
$\Delta T$ is determined by Fourier's law, which reads approximately
$P_\text{abs} / A \sim \kappa\,\Delta T / Z$, where $Z$ is a relevant length
dimension for the test mass (both the thickness and diameter are of similar magnitude for
Cosmic Explorer).  This suggests that in the case of a few watts of laser power
absorbed in the coating (i.e., a coating absorption of roughly \SI{1}{ppm}), the
temperature variation in the substrate should be of order tens of millikelvins,
which is within the $\pm\SI{2.3}{\kelvin}$ limit set by the thermoelastic noise
coupling.

The same thermal fluctuations that drive thermoelastic noise also cause phase
fluctuations in light passing through the substrates, which is relevant for the
two input test masses (ITMs).  For both silica and cryogenic silicon, this phase
fluctuation is dominated by changes in the index of refraction via the
thermorefractive coefficient $\beta = \rmd n / \rmd
T$~\cite{2000PhLA..271..303B,2011PhRvD..84f2001H,2004PhLA..324..345B}.
The power spectrum of this noise is $S(f) \propto \kappa T^2 \beta^2 H /
\mathcal{F} w^4 f^2$, where $H$ is the thickness of the test mass, and
$\mathcal{F} \simeq 2\pi/\mathcal{T}_\text{i}$ is the finesse of the arm cavities, and $\mathcal{T}_\text{i}$ is the transmissivity of the input test masses.  For
fused silica, this noise is well below the other test mass thermal noises.  For
cryogenic silicon, the higher thermal conductivity and larger thermorefractive
coefficient make this noise non-negligible; with the choice of $\mathcal{F}
\simeq 450$, the thermorefractive noise at \SI{10}{\Hz} dominates the total
test mass thermal noise for the \SI{2}{\um} technology,\footnote{The finesse could be increased to decrease the thermorefractive noise and the power absorbed in the input test mass substrates; however, this value is chosen as a compromise to reduce the effects of signal extraction cavity loss on the high frequency quantum noise, which favors small $\mathcal{F}$.} and is similar in
magnitude to the coating Brownian thermal noise at \SI{10}{\Hz} for the
\SI{1}{\um} technology.

Additionally, the semiconductor nature of silicon gives rise to refractive
index fluctuations due to the motion of free carriers in the silicon test
masses.  Initial estimates of this noise source~\cite{P1400084} suggested that
the phase noise induced by these fluctuations could be significant, but a more
recent analysis that includes Debye screening indicates that this noise will
lie several orders of magnitude below the total thermal noise of the
substrate~\cite{P2000048}.  We therefore do not consider this noise source.

Finally, we remark on the static birefringence effects in the test mass substrates.
Cosmic Explorer, like current gravitational-wave laser interferometers, is designed
to operate in a single linear polarization; interconversion of polarization inside
the interferometer acts as an optical loss. The greatest potential for polarization
interconversion is in the substrates of the input test masses, and consequently
the optical gain of the power-recycling cavity could be impacted. Given a mass
thickness of $H$ and a birefringence $\Delta n$, the power-recycling gain is
limited to $G < 1/\sin^2(\pi \Delta n\,H/\lambda)$~\cite{1994OptCo.112..245W};
maintaining $G = 65$ therefore requires $\Delta n \lesssim \num{e-7}$.  This
already appears achievable in existing fused-silica interferometers, and in
laboratory measurements of monocrystalline silicon~\cite{2016CQGra..33a5012K};
for the large-diameter masses of Cosmic Explorer, particularly for the silicon
technology which has not yet been demonstrated for kilometer-scale instruments,
small birefringence must be maintained over a large area, requiring good optical
isotropy and control of the stresses in the substrate.

\subsubsection{Coating noises}
\label{subsubsec:TMthermal_coating}

As with the test mass substrates, the thin-film coatings applied to the test masses also exhibit thermal noises that are driven by mechanical and thermodynamic fluctuations.

The \SI{1}{\um} technology assumes the same target set for the LIGO A+ coatings: an effective factor of 4 overall reduction in mechanical loss compared to the current Advanced LIGO coatings.
This will likely be achieved using silica for the low-index layers, and a yet-to-be-determined metal oxide (or set of metal oxides) for the high-index layers.
Recent measurements indicate that the loss angle of thin-film silica can be as low as \num{2.3e-5}~\cite{2020ApOpt..59A.229G}; to reach the $4\times$ loss reduction target, this requires a loss angle of the high-index layers of \num{7.0e-5}.
The \SI{2}{\um} technology assumes LIGO Voyager coatings, where the low-refractive-index layer is again SiO$_2$, but the high-refractive-index is now amorphous silicon (aSi) with at most \SI{1}{ppm} optical absorption~\cite{2020arXiv200111173A}.

The coating Brownian noise is computed using the formalism of Hong et~al.~\cite{2013PhRvD..87h2001H}, with the photoelastic effect ignored and the loss angle in bulk and shear strains assumed to be equal.\footnote{A formula for Brownian noise under these assumptions is given by Eq.~1 of Yam et~al.~\cite{2015PhRvD..91d2002Y}, but the expression for their coefficient $b_j$ has an error; the corrected expression using their notation is~\cite{2020PhRvL.125a1102T}
\[
    b_j = \frac{1}{1 - \sigma_j} \left[\left(1 - n_j \frac{\partial \phi_\text{c}}{\partial \phi_j}\right)^2 \frac{Y_\text{s}(1 - \sigma_j - 2 \sigma_j^2)}{Y_j(1 - \sigma_\text{s} - 2\sigma_\text{s}^2)}
        + \frac{Y_j(1 - \sigma_\text{s} - 2\sigma_\text{s}^2)}{Y_\text{s}(1 + \sigma_j)}\right].
\]}

As in the substrates, thermodynamic fluctuations produce phase fluctuations of the light propagating in the coatings.
The phase fluctuations are mediated by the coating's average coefficients of thermal expansion $\overline{\alpha}_\text{c}$ and thermorefraction $\overline{\beta}_\text{c}$.
Because of the etalon effect, these coefficients act with opposite sign, leading to an overall thermo-optic effect that for most coatings---including the Cosmic Explorer coatings---is smaller than the thermoelastic or thermorefractive effects individually~\cite{2008PhRvD..78j2003E}.

\subsection{Quantum noise}
\label{subsec:quantum}

The quantum vacuum fluctuations of the modes of the electromagnetic field that enter the antisymmetric port of the interferometer are a significant source of noise at all frequencies~\cite{1981PhRvD..23.1693C, 2002PhRvD..65b2002K, 2002PhRvD..65d2001B}. Quantum radiation pressure noise is caused by the laser light in the arm cavities beating with the vacuum fluctuations in the amplitude quadrature of these modes, producing a fluctuating radiation pressure force acting on the test masses. Shot noise is caused by the beating of the laser with the vacuum fluctuations in the orthogonal phase quadrature, which carries the gravitational wave strain signal.

It is possible to alter the correlations between the fluctuations in these two quadratures in order to modify the quantum radiation pressure and shot noises. This is a rich subject which we do not attempt to review here; see, for example, Refs.~\cite{2012LRR....15....5D, 2014ASSL..404..291H, Miao2012} and references therein. We summarize only those aspects strictly relevant to the Cosmic Explorer design outlined in \cref{sec:limits}. Radiation pressure dominates at low frequencies with a strain power spectral density $S(f) \propto P_\text{arm}/\lambda M^2f^4$. Shot noise dominates at higher frequencies; within the bandwidth of the instrument, the strain power spectral density of the shot noise is $S\propto \lambda/P_\text{arm}$, where $P_\text{arm}$ is the power in the arm cavities and $M$ is the mass of the test masses. The crossover occurs at a frequency $\propto \sqrt{P_\text{arm}/M\lambda}$ of about \SI{10}{\Hz} for Cosmic Explorer. The \SIrange[range-phrase=~and~]{1}{2}{\um} technologies have arm powers of \SIrange[range-phrase=~and~]{1.5}{3}{\mega\W}, respectively, so both realizations of CE2 have the same level of quantum noise.

Squeezed vacuum states~\cite{2017PhR...684....1S, 2019RPPh...82a6905B} can be injected into the antisymmetric port to reduce the noise in one quadrature at the expense of increasing the noise in the orthogonal quadrature, a technique which is being used in Advanced LIGO~\cite{2019PhRvL.123w1107T} and Advanced Virgo~\cite{2019PhRvL.123w1108A}. Therefore, this necessitates a tradeoff between reducing radiation pressure at low frequencies and shot noise at high frequencies. However, the frequency dependence necessary to achieve a broadband noise reduction can be realized by first reflecting the squeezed vacuum off of a detuned optical cavity, known as a filter cavity, before injection into the interferometer~\cite{2002PhRvD..65b2002K, 2013PhRvD..88b2002E, 2014PhRvD..90f2006K}. The production of these frequency dependent squeezed vacuum states has been realized experimentally~\cite{2016PhRvL.116d1102O, 2020PhRvL.124q1102M} and will be used in LIGO A+ and Advanced Virgo+.

Cosmic Explorer will employ a \SI{4}{\km} long filter cavity to achieve a broadband quantum noise reduction of \SI{6}{\decibel} for CE1 and \SI{10}{\decibel} for both realizations of CE2. The filter cavity is critical in achieving the low frequency goals: without it and with this level of squeezing at mid to high frequencies, CE1 would be limited by radiation pressure down to \SI{10}{\Hz} and CE2 would be limited down to \SI{5}{\Hz}.
 
\subsection{Residual gas noise}
\label{subsec:resgas}

The residual gas in the vacuum system is responsible for two noise sources. The first is a phase noise caused by fluctuations of the gas column density in the beam tubes. The contribution to this noise from a particular molecular species with partial pressure $P_\text{tube}$ in the tube, mass $m$, and polarizability $\alpha$ is white up to a cutoff frequency $\propto v_T w_0 / L \lambda$, determined by the time it takes for a molecule to cross the laser beam, with a power spectrum $S\propto \alpha^2 m^{1/2} w_0 P_\text{tube} / T_\text{tube}^{3/2} \lambda$, where $T_\text{tube}$ is the temperature of the tube, $\lambda$ is the wavelength, $w_0$ is the laser beam's waist, $L$ is the length of the arm, and $v_T$ is the thermal velocity of the molecule.~\cite{1996magr.meet.1434Z,2002JVST...20.1237T}.

The second is a force noise caused by the residual gas in the chambers exerting a damping force on the test masses. The contribution of one molecular species with partial pressure $P_\text{chamber}$ in a chamber to this noise has a power spectrum $S(f)\propto T_\text{chamber}^{1/2} m^{1/2} R^2 P_\text{chamber} / M^2 f^4$, where $M$ and $R$ are the mass and radius of the test mass, respectively, and $T_\text{chamber}$ is the temperature of the chamber~\cite{2010PhLA..374.3365C,2011PhRvD..84f3007D}. The magnitude of these two noise sources determine the pressure requirements in the beam tubes and test mass chambers for each gas species described in \cref{subsec:facility_tube}.
 
\subsection{Scattered light noise}
\label{subsec:scatter}

\begin{figure}
  \centering
  \includegraphics[width=\figwidth]{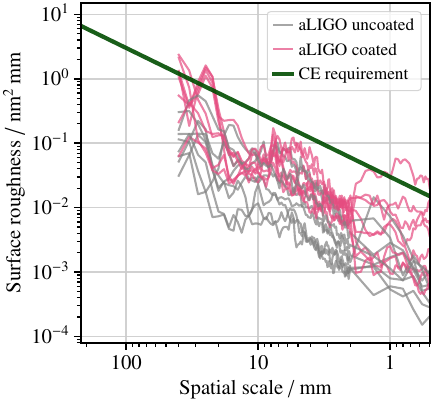}
  \caption{Requirement on surface roughness used to calculate small angle scattering shown in \cref{fig:scatter}, along with the measured spectra from Advanced LIGO test masses. Due to Cosmic Explorer's large beam sizes, the relevant spatial scale (inverse spatial frequency) of the mirror roughness extends to several tens of centimeters.}
  \label{fig:mirror_psd}
\end{figure}

\begin{figure*}
    \centering
    \includegraphics[width=\figwidth]{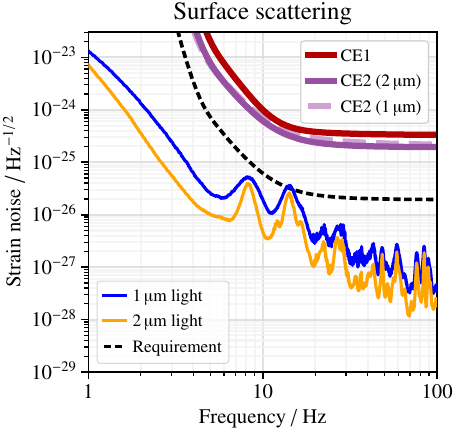}
    \includegraphics[width=\figwidth]{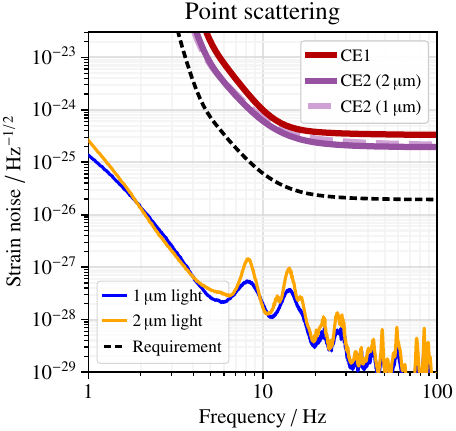}
    \caption{Back-scattering noise for surface roughness (left) and point defects (right) for a \SI{120}{\cm} diameter beam tube with \SI{100}{\cm} diameter baffle apertures. The black dashed curve shows the facility requirement that the scattering noise be a factor of ten below the minimum of the three design noise curves shown for Cosmic Explorer. The BRDF for surface roughness scattering is proportional to the target PSD shown in \cref{fig:mirror_psd} and the point scattering BRDF is \SI{e-4}{\per\steradian}. The BRDF of the baffles and beam tube is \SI{e-3}{\per\steradian}. The peaks are due to beam tube resonances.}
    \label{fig:scatter}
\end{figure*}

Scattering of light within the beam tubes is a source of noise for all ground-based interferometric gravitational wave detectors, as first calculated by Thorne~\cite{thorne_s}. Imperfections on the surface of the test masses lead to scattering of the main cavity mode, which can be broadly grouped into two classes:
\begin{itemize}
    \item
    \textit{surface roughness}, which are variations on the test mass surface responsible mostly for scattering at narrow angles; and
    \item
    \textit{point defects}, which are ``bright spots'' on the mirror's surface that produce diffuse scattering and are therefore responsible mostly for scattering at wide angles.
\end{itemize}
These imperfections on the test masses cause light to scatter out of the cavity and reflect multiple times off the beam tube wall as it propagates down the tube, and eventually recombine with the main cavity mode at the opposite test mass. Seismic motion of the beam tube imposes a phase noise on the scattered light each time it reflects off the tube, and gives rise to readout noise when the light recombines. Scattering of this nature was first pointed out by Thorne as an important noise source for the LIGO beam tubes (see Section III.B. of \cite{thorne_s}). To address this, baffles were installed at various points along the LIGO beam tubes to deflect scattered light away from the test masses. However, the baffles give rise to \textit{back-scattering noise}, whereby light that is scattered out of the cavity by one of the test masses is back-scattered off one of the baffles and subsequently recombines with the main cavity mode at the same test mass. Motion of the beam tube then imposes a phase noise on the back-scattered light, which gives rise to readout noise when the light recombines. A detailed explanation of this effect is given by Flanagan and Thorne~\cite{flanagan_thorne_bs} and a detailed analysis of back-scattering specifically for Cosmic Explorer is given in a recent technical report~\cite{bai_scatter}, which we summarize below. The effect of forward scattering, meaning the diffraction of the main beam whose time dependence arises primarily from seismically induced transverse motion of the baffles, is left for future work~\cite{Flanagan1995,2001JOSAA..18..546N}. Additionally, the phase information of the baffle surfaces is not considered in this work, though simulations on Advanced LIGO indicate that the inclusion of this information can cause the scatter-induced strain noise power spectral density to fluctuate by an order of magnitude in either direction~\cite{Day2013}.

The fractional power scattered per unit solid angle is quantified by the bidirectional reflectance distribution function (BRDF), and the power spectrum of noise due to back scattered light is $S\propto \beta\kappa S_\xi$, where $\beta$ is the BRDF of the backscattering surface (i.e., the baffles and beam tube) and $\kappa$ is related to the square of the mirror BRDF. Furthermore, $S_\xi$ is the longitudinal displacement noise of the beam tube, taking into account fringe-wrapping as explained in~\cite{2012OExpr..20.8329O} and Section 3.1 of~\cite{bai_scatter}. Here we use beam tube motion measured from the LIGO Livingston observatory, but the Cosmic Explorer baffles can be suspended to reduce their motion if necessary. Cosmic Explorer will likely use baffles with a black nickel coating with a BRDF of \SI{e-3}{\per\steradian}~\cite{Ananyeva2019}, however diamond-like carbon coatings with a BRDF of \SI{e-4}{\per\steradian} can be used if necessary.

Surface roughness of spatial frequency $\nu$ gives rise to scattering at angle $\theta \sim \lambda \nu$, where $\lambda$ is the optical wavelength, and $\theta$ is measured relative to the beam tube axis. The BRDF for this small angle scattering due to surface roughness is proportional to the PSD of the mirror surface variations at spatial frequency $\nu$. The left panel of \cref{fig:scatter} shows the noise due to surface scattering using the surface PSD shown in \cref{fig:mirror_psd}, assuming a \SI{120}{\cm} tube diameter and \SI{100}{\cm} baffle aperture diameter. This PSD, with functional form $S(\nu) = (\SI{0.03}{\nm^2\,\mm}) / (\SI{1}{\mm} \times \nu)$, is an upper limit requirement on the surface roughness over an appropriate range of spatial scales that scatter into the tube, based on Sec.~2.2 of~\cite{bai_scatter}, which results in noise due to surface scattering that is at least a factor of ten below the design sensitivity at all frequencies. This requirement is comparable to the surface roughness that has already been achieved with the Advanced LIGO test masses at spatial scales below a few centimeters. (Comparison is harder at larger spatial scales, where the Advanced LIGO surface roughness is not well characterized.)

Point defects give rise to diffuse scattering, which has a roughly constant BRDF.
The right panel of \cref{fig:scatter} shows the scattered light noise due to point defects assuming a mirror BRDF of \SI{e-4}{\per\steradian}, and a \SI{120}{\cm} tube diameter. It appears that diffuse scattering is an insignificant noise source for all phases of Cosmic Explorer.

The choice of beam tube diameter is of particular importance for the design of Cosmic Explorer. While wider tubes lead to less scattering noise, they are also substantially more expensive considering the cost of the vacuum envelope, the metal needed for the tube, and the market availability of various tube dimensions. \cref{fig:scatter} shows that a \SI{120}{\cm} diameter tube is sufficient to keep the back-scattering noise below the noise requirements for both phases of the interferometer, provided that the requirements on mirror surface roughness are met. This limit on beam tube size will be reevaluated once the effects of forward scattering are considered.
 
\subsection{Noise associated with controls}
\label{subsec:controls}

As a practical matter, the relative distances between the suspended optics as well as their angular alignment must be precisely servo controlled in order to keep the interferometer stable and operating in the linear regime. Noise from the sensors used to measure the linear and angular degrees of freedom is imposed on the optics by the control systems needed to suppress their relative positions and orientations.

In addition to the differential arm motion of the four test masses, there are three auxiliary length degrees of freedom of the other core optics, which are suspended from triple pendulum suspensions, that must be controlled. These degrees of freedom are limited by seismic noise below a few hertz and by sensing noise (of similar magnitude to that of Advanced LIGO) at higher frequencies. The auxiliary degree of freedom with the strongest coupling to the differential arm motion is the Michelson degree of freedom: differential motion between the beamsplitter and the input test masses also produce phase fluctuations at the anti-symmetric port. The Michelson degree of freedom is suppressed by a factor of $\pi/2\mathcal{F}\simeq \num{3.5e-3}$ relative to the differential arm motion since the latter is enhanced by the Fabry-P\'{e}rot arm cavities. The Michelson sensing noise is of order \SI{e-16}{\m\,\Hz^{-1/2}}~\cite{2016PhRvD..93k2004M}, which gives an equivalent strain sensitivity of $\sim\SI{6e-24}{\Hz^{-1/2}}$. Simulations show that if this motion is sensed and subtracted from the differential arm motion, a control loop with a bandwidth of a few hertz is sufficient to suppress the Michelson noise to within a factor of 10 below the design sensitivities for both CE1 and CE2. Simulations also suggest that the couplings of the other two auxiliary length degrees of freedom, fluctuations in the power recycling and signal extraction cavity lengths, do not significantly couple to the differential arm motion through the fundamental optical mode.

The noise from the angular control systems is one of the most challenging low frequency technical noise sources in current gravitational wave detectors, and it is also expected to be so for third generation detectors. Radiation pressure from the circulating arm power exerts a torque on the mirrors. This torque stiffens (or hardens) the torsional resonance when the cavity mirrors rotate with the same sign, and softens the resonance when the mirrors rotate with opposite sign~\cite{2006PhLA..354..167S,2010CQGra..27h4026B,2010ApOpt..49.3474H,2013JOSAA..30.2618D}. The hard and soft resonances are shifted by $\Delta f_\text{h,s}^2 = \gamma_\text{h,s} P_\text{arm}L_\text{arm}/Ic$, where $I$ is the moment of inertia of the mirrors, $\gamma_\text{h} > 0$ is a geometric factor for the hard mode, and $\gamma_\text{s} < 0$ is a geometric factor for the soft mode. The soft mode will become unstable if the torque is large enough and the (negative) shift $\Delta f_\text{s}^2$ exceeds the mechanical resonance $f_0^2$. In this case, the bandwidth of the angular control loop needs to be several times the frequency of this unstable mode in order to stabilize the optomechanical system. Achieving this requirement without injecting excess sensing noise is challenging.

It is thus clearly advantageous to prevent the soft mode from becoming unstable. In this case the control loop bandwidth needs to be $\sim 3f_\text{s}$~\cite{2020arXiv200111173A}. One way to achieve this is to reduce the frequency shift $\Delta f_\text{s}^2$. The arm power and length are set and the geometric factor is constrained by the necessity of minimizing the beam spot sizes. However, the moment of inertia can be increased, perhaps by increasing the test mass thickness or altering the geometry in some other way. Another possibility is to increase the free torsional resonance $f_0$. The soft mode frequency shifts $\Delta f_\text{s}^2$ are approximately $-(\SI{0.6}{\Hz})^2$ for the \SI{1}{\um} technology and $-(\SI{1.0}{\Hz})^2$ for the \SI{2}{\um} technology. The soft mode will thus be stable, necessitating a sufficiently low loop bandwidth of a few hertz, if $f_0\gtrsim \SI{1}{\Hz}$.

Even though the frequency shift $\Delta f_\text{h}^2$ for the hard mode is always positive and the hard mode always stable, it can still be excited and must be damped. Two factors make this requirement intrinsically easier for Cosmic Explorer than for Advanced LIGO. First, the typical amplitude of these excitations will be less due to the improved seismic isolation. Second, the geometric factor for the hard mode is, to first order, proportional to $(w/w_0)^4$ where $w_0$ and $w$ are the beam radii at the waist and at an optic, respectively~\cite{2006PhLA..354..167S}. The ratio $w/w_0$ needs to be small for CE to reduce diffraction over \SI{40}{\km}, while for Advanced LIGO it is made large to reduce coating thermal noise. This results in hard mode frequency shifts $\Delta f_\text{h}^2$ of approximately $+(\SI{1.1}{\Hz})^2$ for the \SI{1}{\um} technology and $+(\SI{2.1}{\Hz})^2$ for the \SI{2}{\um} technology.

We have only sketched the requirements for the control system and its noise performance here; while these preliminary considerations suggest that it will be possible to meet the low frequency requirements, a realistic understanding of the control noise is a significant source of uncertainty facing Cosmic Explorer and warrants a more detailed analysis.
 
\section{Discussion and Conclusion}
\label{sec:discussion}

\begin{table*}
  \centering
  \begin{tabular}{p{23em} l l c c c c c}
    \toprule
    Activity &
    Theme   &
    Fac.    &
    CE1     &
    CE2(1)  &
    CE2(2)
    \\
    \hline

Partial pressures of gases (\ref{subsec:facility_tube}) &
    Vacuum &
    $\bullet$ &  &  &
    \\
Ambient seismic field characterization, incl.\ surface and body wave content (\ref{subsec:facility_seismic}) &
    Seismic arrays &
    $\bullet$ &  &  &
    \\
Ambient infrasound field characterization, distinguished from wind-induced sensor noise (\ref{subsec:facility_atmospheric}) &
    Infrasonic arrays &
    $\bullet$ &  &  &
    \\
Reduction of seismic field near test masses (\ref{subsubsec:newtonian_mitigation}) &
    Seismic metamaterials &
    $\bullet$ &  &  &
    \\
Reduction of magnetic field coupling &
    Other environmental &
    $\bullet$ &  &  & 
    \\
\hline
\SI{1}{\pico\meter\, \Hz^{-1/2}} horiz.\ susp.\ point motion at \SI{1}{\Hz} (\ref{subsec:seismic}) &
    Inertial sensing &
      & $\bullet$ &  &
    \\
2$\times$ subtraction of surface-wave NN (\ref{subsubsec:newtonian_seismic}) &
    Seismic arrays &
      & $\bullet$ &  &
    \\
\SI{1.5}{\mega\watt} \SI{1}{\um} arm power and \SI{6}{\decibel} FD squeezing (silica) &
    QN, scatter &
      & $\bullet$ &  &
    \\
Silica test mass, \SI{70}{\cm} $\varnothing$; low impurity &
    Silica materials science &
      & $\bullet$ & $\bullet$ &
    \\
Highly stressed silica blade springs (\ref{subsec:susthermal}) &
    Silica materials science &
      & $\bullet$ & $\bullet$ &
    \\
Validation of silica loss mechanisms at \SI{5}{\Hz}  &
    Silica materials science &
      & $\bullet$ & $\bullet$ &
    \\
A+ coatings over \SI{70}{\cm} $\varnothing$ (\ref{subsubsec:TMthermal_coating}) &
    Thin-film mirror coatings &
      & $\bullet$ & $\bullet$ &  &
    \\
Test mass surface polishing of large substrates (\ref{subsec:scatter}) &
    Mirror metrology &
      & $\bullet$ & $\bullet$ & $\bullet$
    \\
Control noise  &
    Optical sensing and control &
      & $\bullet$ & $\bullet$ & $\bullet$
    \\
\hline
10$\times$ subtraction of surface-wave NN (\ref{subsubsec:newtonian_seismic}) &
    Seismic arrays &
      &  & $\bullet$ & $\bullet$
    \\
3$\times$ subtraction of body-wave NN (\ref{subsubsec:newtonian_seismic}) &
    Seismic arrays &
      &  & $\bullet$ & $\bullet$
    \\
Best effort at mitigation of infrasonic NN &
    Infrasonic arrays &
    $\bullet$ &  & $\bullet$ & $\bullet$
    \\
\SI{0.1}{\pico\meter\, Hz^{-1/2}} horiz.\ susp.\ point motion at \SI{1}{\Hz} (\ref{subsec:seismic}) &
    Inertial sensing &
      &  & $\bullet$ & $\bullet$
    \\
\SI{1.5}{\mega\watt} \SI{1}{\um} arm power and \SI{10}{\decibel} FD squeezing (silica) &
    QN, scatter &
      &  & $\bullet$ &
    \\
\hline
Silicon test mass, \SI{80}{\cm} $\varnothing$; low impurity &
    Silicon materials science &
      &  &  & $\bullet$
    \\
Highly stressed silicon blade springs and ribbons (\ref{subsec:susthermal}) &
    Silicon materials science &
      &  &  & $\bullet$
    \\
Validation of silicon loss mechanims at \SI{5}{\Hz}  &
    Silicon materials science &
      &  &  & $\bullet$
    \\
``Voyager'' coatings over \SI{80}{\cm} $\varnothing$ (\ref{subsubsec:TMthermal_coating}) &
    Thin-film mirror coatings &
      &  &  & $\bullet$
    \\
\SI{3.0}{\mega\watt} \SI{2}{\um} arm power and \SI{10}{\decibel} FD squeezing (silicon) &
    QN, scatter &
      &  &  & $\bullet$
    \\
Radiative temperature control to $\pm$\SI{2}{\kelvin} (\ref{subsubsec:TMthermal_substrate}) &
    Cryogenics &
    $\bullet$ &  &  & $\bullet$
    \\
    \botrule
  \end{tabular}
  \caption{Summary of required research and development activities.
    The final columns in the table indicate whether the activity involves primarily the facility, the initial Cosmic Explorer detector (CE1), or the advanced Cosmic Explorer detector (CE2); for the advanced detector, activities are presented for both the scenario in which the detector is room-temperature glass technology with \SI{1}{\um} lasers, or cryogenic silicon technology with \SI{2}{\um} lasers.}
  \label{tab:research}
\end{table*}
 
In this work we have presented updated sensitivity curves for Cosmic Explorer
and have also identified several areas of research and development that will be
necessary to realize its low-frequency performance.
\begin{itemize}
    \item the identification of a facility site with low seismic and acoustic noise, and other suitable environmental properties;
    \item the development of low-noise inertial isolators in multiple degrees of freedom;
    \item the continued development of mitigation techniques for Newtonian noise;
    \item the production of large, high-quality test mass substrates, both silica and silicon;
    \item the polishing and coating of large test mass substrates to a resulting spatial roughness comparable to that achieved for the Advanced LIGO test masses, but characterized at larger spatial scales;
    \item the development of suitable mirror coatings;
    \item the development of long multi-stage suspensions employing highly stressed silica and silicon blade springs and silica fibers and silicon ribbons to support \SI{320}{\kg} test masses;
    \item the development of alternatives to blade spring suspensions, such as geometric anti-springs;
    \item the validation and extension of the beam-tube scattering model presented here;
    \item the development of a robust angular control system with possible modifications to the suspensions and/or test masses to reduce the effects of radiation pressure instabilities;
    \item the development of vacuum technology and practice capable of achieving ultra-high vacuum in both the test mass chambers, which will be periodically vented, and the beam tubes;
    \item the measurement of material properties, such as mechanical loss angles, down to \SI{5}{\Hz};
and
    \item the development of laser frequency and intensity noise requirements and the optical topologies required to achieve them not discussed here.
\end{itemize}
\cref{tab:research} summarizes the research required to reach the low frequency sensitivity presented here along with a rough timeline of when that research would need to be completed.

We have also shown that both the \SI{1}{\um} and \SI{2}{\um} technologies can
realize nearly identical low frequency sensitivities for CE2. While this is
true for high frequencies as well, achieving the specified quantum and thermal
noise performance for both technologies requires further research and
development not discussed in this paper.  Additionally, if the arm length of
Cosmic Explorer were significantly shortened, the relative importance of the various noise sources may change since they scale differently with arm length~\cite{2017CQGra..34d4001A}.

\begin{acknowledgments}
The authors gratefully acknowledge the support of the National Science Foundation through collaborative award numbers 
1836814, 1836809, 1836734, and 1836702. EDH is supported by the MathWorks, Inc.
JRS is partially supported by the Dan Black Family Trust.
BK is supported by the Heising-Simons Foundation.
This work has document numbers CE--P2000005 and LIGO--P2000405.
\end{acknowledgments}

\appendix
\section{Summary of Cosmic Explorer Technologies}
\label{sec:technologies}

\begin{figure}[t]
  \centering
  \includegraphics[width=\figwidth]{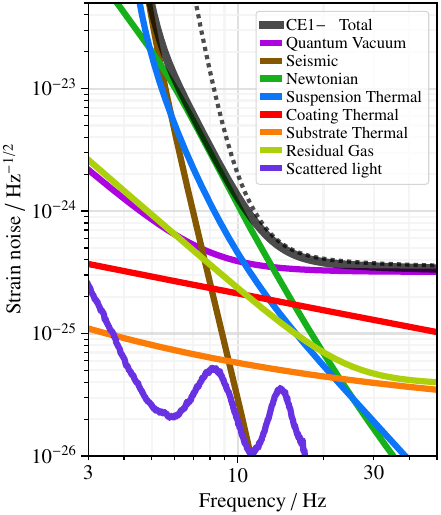}
  \caption{
    Estimated low-frequency spectral sensitivity limit (solid black) of Cosmic Explorer 1$-$ and the known noise sources that cause these limits (colored curves).
    The sensitivity limit for Cosmic Explorer 1 from previous work~\cite{2017CQGra..34d4001A} is also shown (dotted black curve).}
  \label{fig:CE1minus}
\end{figure}

One advantage of realizing Cosmic Explorer incrementally is that CE1 can achieve significantly higher sensitivities than the second generation detectors mostly using the existing technology developed for LIGO A+. In addition to providing a relatively short route to increased sensitivity, this provides some risk management: significant improvements can still be made even if some advanced technologies are not realized. Nevertheless, the baseline CE1 design does rely on some technological advances beyond A+. We can also consider a more conservative detector, CE1$-$, which relies solely on A+ technology with the improved sensitivity coming only from scaling up the arm length, test masses, and suspensions from the A+ design. In particular, this would differ from CE1 by the following:
\begin{itemize}
    \item No fused silica blade springs on the final suspension stage between the PUM and the test mass. The suspensions are just a scaled up version of the A+ suspensions.
    \item No Newtonian Rayleigh wave suppression.
    \item The same level of suspension point motion as A+, a factor of 10 worse than CE1 at \SI{1}{\Hz}.
\end{itemize}
\cref{fig:CE1minus} shows the low-frequency limit to the spectral sensitivity of CE1$-$. These changes only affect the low-frequency noise below about \SI{20}{\Hz} leaving identical high-frequency sensitivities for CE1 and CE1$-$. This can also be thought of as an initial detector to be implemented first while some of the above technologies are being developed for CE1 if necessary.

\begin{figure}[t]
  \centering
  \includegraphics[width=\figwidth]{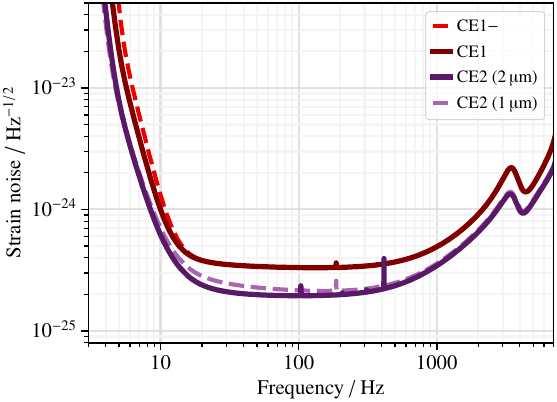}
  \caption{Strain sensitivities of the different Cosmic Explorer technologies and detectors.}
  \label{fig:stages}
\end{figure}

A summary of the defining parameters of the different Cosmic Explorer detectors and technologies is given in \cref{tab:tech_params} and their sensitivities compared in \cref{fig:stages}; many of the other details common to all detectors using the same technology are given in \cref{tab:tm_params}. All of the \SI{1}{\um} detectors share the same basic properties: arm power, material, temperature, coatings, and beam spot sizes. The low-frequency sensitivity of CE1 is improved over CE1$-$ by the addition of fused silica blades springs, which reduce the suspension thermal and seismic noises as described in \cref{subsec:susthermal,subsec:seismic}; improved seismic isolation, as discussed in \cref{subsec:seismic}; and some suppression of Newtonian Rayleigh waves, as discussed in \cref{subsec:newtonian}. The test mass thermal noises, most importantly coating Brownian, are the same for all detectors using \SI{1}{\um} technology since they use the same test mass substrates and coatings, beam sizes, and temperatures.

\begin{table}[bh]
  \centering
  \sisetup{
    table-format=3.1,
    table-space-text-post=\hspace{-1em},
    }
  \begin{tabular}{r s S S S S}
    \toprule
    \textbf{Quantity} & \textbf{Units} & \textbf{CE1$-$} & {\textbf{CE1}} & {\textbf{CE2 (\SI{1}{\um})}} & {\textbf{CE2 (\SI{2}{\um})}} \\
    \hline
    Arm power & \mega\watt & 1.5 & 1.5 & 1.5 & 3 \\

    Wavelength & \um & 1 & 1 & 1 & 2 \\

    Squeezing & \decibel & 6 & 6 & 10 & 10 \\

    Material & & {Silica} & {Silica} & {Silica} & {Silicon} \\

    Temperature &
    \K &
    293 &
    293 &
    293 &
    123 \\

    Final stage blade & & {No} & {Yes} & {Yes} & {Yes} \\

    Rayleigh wave suppr. & & {None} & {$2\times$} & {$10\times$} & {$10\times$} \\

    Body wave suppr. & & {None} & {None} & {$3\times$} & {$3\times$} \\

    Susp. point at \SI{1}{\Hz} & \si{\pico\meter\,\Hz^{-1/2}} & 10 & 1 & 0.1 & 0.1 \\

    Coatings & & {A+} & {A+} & {A+} & {Voyager} \\

    ITM spot size &
    \cm &
    \sisetup{round-mode=figures,round-precision=2}12.0 &
    \sisetup{round-mode=figures,round-precision=2}12.0 &
    \sisetup{round-mode=figures,round-precision=2}12.0 &
    \sisetup{round-mode=figures,round-precision=2}16.4 \\

    ETM spot size &
    \cm &
    \sisetup{round-mode=figures,round-precision=2}12.0 &
    \sisetup{round-mode=figures,round-precision=2}12.0 &
    \sisetup{round-mode=figures,round-precision=2}12.0 &
    \sisetup{round-mode=figures,round-precision=2}16.4 \\

\botrule
  \end{tabular}
  \caption{Defining parameters of the different Cosmic Explorer technologies and detectors. See \cref{tab:tm_params} for more details common to all detectors using the same technology.}
\label{tab:tech_params}
\end{table}
 
The high frequency sensitivity of CE2 is nearly identical for both the \SI{1}{\um} and \SI{2}{\um} technologies since this is determined by quantum shot noise. The \SI{1}{\um} realization of CE2 has the same squeezing as the \SI{2}{\um} realization---\SI{10}{\decibel} increased from \SI{6}{\decibel} for CE1. Since the shot noise scales as $S\propto \lambda/P_\text{arm}$, the factor of two larger power stored in the arms of the \SI{2}{\um} realization gives the same shot noise level as the \SI{1}{\um} realization. All other technologies not dependent on test mass material or laser wavelength are the same for both realizations of CE2. In particular, the seismic isolation is improved over that of CE1 by a factor of 10 at \SI{1}{\Hz}, Newtonian body waves are suppressed by a factor of three, and Newtonian Rayleigh waves are suppressed by an additional factor of five over that of CE1. Both realizations thus have the same Newtonian noise.

To summarize, the low frequency sensitivity is dominated by suspension thermal, seismic, and Newtonian noise. The low frequency sensitivity of CE1 is improved over that of CE1$-$ due to improved suspensions, seismic isolation, and the addition of Newtonian noise suppression. The low frequency sensitivity of CE2 is improved over that of CE1 through further improvements to the seismic isolation and Newtonian noise suppression and increased squeezing. Since the high frequency sensitivity is determined by quantum shot noise, CE1 and CE1$-$ have the same high frequency sensitivity, as do both realizations of CE2.

 \section{Displacement and Force Sensitivity}
\label{sec:displacement_force}

\begin{figure}[t]
  \centering
  \includegraphics[width=\figwidth]{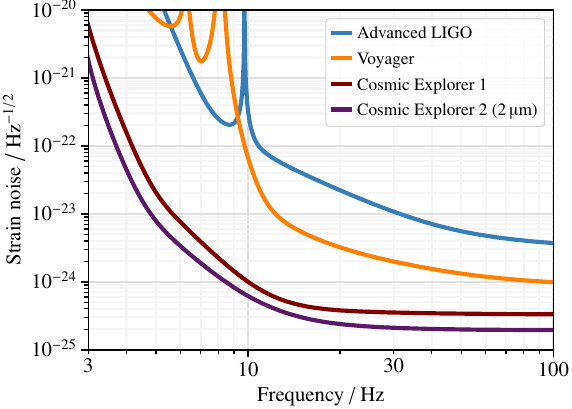}
  \includegraphics[width=\figwidth]{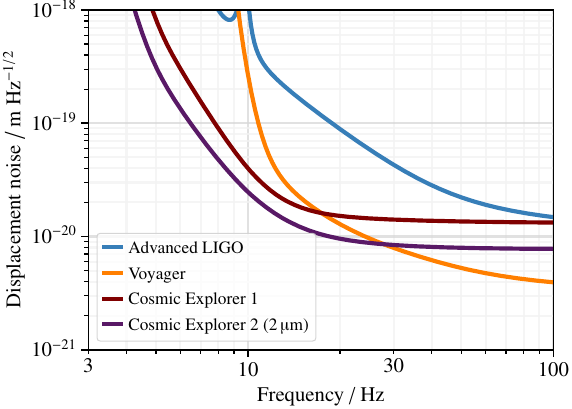}
  \includegraphics[width=\figwidth]{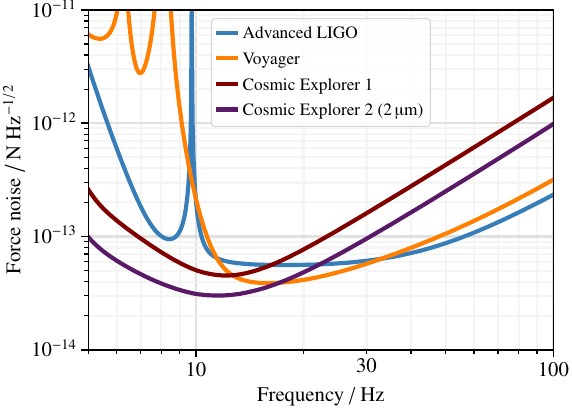}
  \caption{Comparison of Cosmic Explorer strain, displacement, and force noises with those of Advanced LIGO and Voyager.}
  \label{fig:strain_displacement_force}
\end{figure}

\cref{fig:strain_displacement_force} compares the noise of Cosmic Explorer, Advanced LIGO, and Voyager in terms of gravitational wave strain and the equivalent test mass displacement and force noises. To achieve its design sensitivity above ${\sim}\SI{20}{\Hz}$, Cosmic Explorer does not require as low displacement or force noise as does Voyager, owing to the longer arms and larger test masses. However, significant improvements in displacement and force noises are required to achieve the Cosmic Explorer strain sensitivity at lower frequencies.
 
\bibliographystyle{bibtex/bst/revtex/apsrev4-2}
\bibliography{refs_ads,refs_other}

\end{document}